\documentclass[altaffilletter,aps,nofootinbib,twocolumn,prd,eqsecnum,preprintnumbers,superscriptaddress,10pt,floatfix]{revtex4-1}
\pdfoutput=1
\usepackage[caption=false]{subfig}
\usepackage{graphicx}
\usepackage{amsmath}
\usepackage{amsfonts}
\usepackage{mathtools}
\usepackage{amssymb}
\usepackage{enumerate}
\usepackage{xcolor}
\usepackage{bm}
\usepackage{mathrsfs}
\usepackage{epstopdf}
\usepackage{url}
\usepackage{footnote}
\usepackage{textcomp}
\usepackage{dsfont}
\usepackage{ulem}
\usepackage{hyperref}
\usepackage{enumerate}   
\usepackage{appendix}
\usepackage{textcomp}
\usepackage{tipa}

\makeatletter
\newcommand*{\rom}[1]{\expandafter\@slowromancap\romannumeral #1@}
\makeatother

\begin{document}

\title{Quasinormal modes of black holes encircled by a gravitating thin disk}

\author{Che-Yu Chen}
\email{b97202056@gmail.com}
\affiliation{RIKEN iTHEMS, Wako, Saitama 351-0198, Japan}
\affiliation{Institute of Physics, Academia Sinica, Taipei 11529, Taiwan}

\author{Petr Kotlařík}
\email{kotlarik.petr@gmail.com}
\affiliation{Institute of Theoretical Physics, Faculty of Mathematics and Physics, Charles University, V Holešovičkách 2, 180 00 Prague 8, Czech Republic}

\begin{abstract}
The ringdown phase of gravitational waves emitted by a perturbed black hole is described by a superposition of exponentially decaying sinusoidal modes, called quasinormal modes (QNMs), whose frequencies depend only on the property of the black hole geometry. The extraction of QNM frequencies of an isolated black hole would allow for testing how well the black hole is described by general relativity. However, astrophysical black holes are not isolated. It remains unclear whether the extra matter surrounding the black holes such as accretion disks would affect the validity of the black hole spectroscopy when the gravitational effects of the disks are taken into account. In this paper, we study the QNMs of a Schwarzschild black hole superposed with a gravitating thin disk. Considering up to the first order of the mass ratio between the disk and the black hole, we find that the existence of the disk would decrease the oscillating frequency and the decay rate. In addition, within the parameter space where the disk model can be regarded as physical, there seems to be a universal relation that the QNM frequencies tend to obey. The relation, if it holds generically, would assist in disentangling the QNM shifts caused by the disk contributions from those induced by other putative effects beyond general relativity. The QNMs in the eikonal limit, as well as their correspondence with bound photon orbits in this model, are briefly discussed.   
\end{abstract}

\maketitle

\section{Introduction}

Black holes ring when they are perturbed, with the ringing frequencies determined by the underlying spacetime geometry. The feature of the ringings of black holes is tightly related to the fact that the whole system is dissipative. For an asymptotically flat black-hole spacetime, the emitted gravitational waves propagate outward, escaping from the system to spatial infinity. In addition, the event horizon, i.e., a point beyond which no infalling matter can return, acts as the other boundary of dissipation of the system. Because of the dissipation, the ringings of black holes would decay. Such a ``ringdown" phase can be described by a superposition of exponentially decaying sinusoidal oscillations, called quasinormal modes (QNMs) \cite{Kokkotas:1999bd,Berti:2009kk,Konoplya:2011qq}. The QNM frequencies are complex-valued, with the real part describing the oscillations, and the imaginary part determining the decay of the amplitudes. Importantly, for an isolated black hole in general relativity (GR), the spacetime geometry dictates the QNM spectrum, and they both satisfy the no-hair theorem, i.e., they are purely determined by the mass and the spin of the black hole. Therefore, based on the current achievements \cite{LIGOScientific:2016aoc,LIGOScientific:2021djp} and with the upcoming advancements in the gravitational wave detection of binary merger events \cite{Reitze:2019iox,Maggiore:2019uih}, the extraction of QNM frequencies from ringdown signals may be accessible, helping us to identify the black hole parameters and even to test GR.

However, astrophysical black holes are generally not isolated. They may be surrounded by dark matter halos, or be encircled by accretion disks. The validity of using black hole QNMs to extract parameters describing the black hole spacetime requires a sufficient understanding of how the surrounding matter would alter the QNMs. One has to ensure that the contributions from the environments can be disentangled from those induced by the black hole geometry itself, at least under suitable approximations. 

The QNM spectra of black holes surrounded by matter -- the dirty black holes -- have been explored in the literature. In Refs.~\cite{Leung:1997was,Leung:1999iq}, the surrounding matter was modeled by a spherical dust thin-shell. It was shown, both numerically and analytically, that the QNM spectrum could deviate significantly from the vacuum case, especially when the shell is far away from the black hole. It was later clearly elucidated in Refs.~\cite{Barausse:2014tra,Barausse:2014pra}, assuming again spherically symmetric matter configurations, that this large amount of frequency shifts could be actually due to the existence of the double-barrier structure on the effective potential in the QNM master equations. The surrounding matter induces an additional barrier in the effective potential, which could induce pseudospectral instability of black hole QNMs \cite{Jaramillo:2020tuu}. Even an additional tiny bump on the effective potential would already trigger the instability and excite additional modes. The instability could happen even to the fundamental modes -- the longest-lived modes \cite{Cheung:2021bol}, but their frequencies may still be extracted robustly from the prompt ringdown signals in time-domain \cite{Berti:2022xfj} (see also \cite{Kyutoku:2022gbr}). The pseudospectral instability can be avoided when the contributions of the surrounding matter on the effective potential are sufficiently mild in the sense that, in the case of nonrotating black holes, the effective potential retains its single-peak structure. This can be achieved as shown by the model of Ref.~\cite{Cardoso:2021wlq}, in which the authors, assuming spherical symmetry, proposed an effective metric that can describe the spacetime geometry of a whole galaxy harboring a supermassive black hole. In this case, the frequencies of fundamental modes are shifted mildly by the environmental effects. The highly damped QNMs of spherically symmetric dirty black holes were also studied \cite{Medved:2003rga}.          

Apparently, the discussion of the QNM spectrum for dirty black holes so far is still quite confined to the assumption that the overall spacetime remains spherically symmetric. However, in a more realistic scenario, such as a black hole encircled by a gravitating accretion disk, the spherical symmetry is no longer preserved. But, the complicated structure of the Einstein equations makes obtaining the common gravitational field of the black hole with the disk a rather difficult task, at least for analytical work. Some reasonable simplifications (symmetries) are still needed. The simplest viable option is to consider an axially symmetric disk and neglect (or compensate) the total rotation present in the spacetime, so the spacetime is also static, and the black hole is described by the Schwarzschild metric. Another assumption that can be made is that the typical thickness of the disk is much smaller than the black hole radius, thus it is effectively infinitesimally thin. Then the Einstein equations are simplified considerably. Nevertheless, not many models of the Schwarzschild black hole encircled by a thin disk (SBH-disk models) are known in the literature. The first ``superposition'' was made in Ref.~\cite{Lemos:1994} (further studied in Ref.~\cite{Semerak:2000}) using inverted Morgan-Morgan disk \cite{Morgan:1969}. It was also used to calculate the influence of a heavy accretion disk on the black-hole shadow in the more recent work \cite{Cunha:2020}. Another class of disk solutions was proposed in Ref.~\cite{Semerak:2004}, revisited recently in Ref.~\cite{Kotlarik:2022}. Both of these models have a slight disadvantage in that only a part of the metric was obtained explicitly, the rest being left to numerical treatments when needed. Yet recently, new solutions have been found \cite{Vieira:2020,Kotlarik:2022spo}, where the whole metric of the entire superposition was derived explicitly and in closed-forms. In this paper, we consider the SBH-disk model proposed in Ref.~\cite{Kotlarik:2022spo}.

From the astrophysical point of view, the SBH-disk model \cite{Kotlarik:2022spo} may not properly describe any realistic scenario of accretion processes. In addition, being static, it does not include the spin of the central black hole nor the rotation of the disk. However, the disk possesses physically reasonable properties, and it can demonstrate the effects that may actually occur in the real astrophysical setup where the gravitation from the disk cannot be totally neglected.

In the presence of the gravitating disk, the calculations of the QNMs for the SBH-disk model become substantially challenging because the master equations in general are nontrivial partial differential equations. This is true even for the calculations of the QNMs  of massless scalar fields. To proceed, we assume that the mass of the disk is much smaller than the black hole one. Up to the first order of the mass ratio, we adopt the projection method, which was proposed in Ref.~\cite{Cano:2020cao} then applied in Refs.~\cite{Cardoso:2021qqu,Chen:2022ynz,Zhao:2023uam,Ghosh:2023etd}, to derive the master equation and investigate how the QNM frequencies of a massless scalar field are shifted by the gravitating disk. To ensure the validity of the projection method and the stability of the disk, we can fairly consider the parameter space of the model in which the aforementioned pseudospectral instability of fundamental modes does not happen. This can be achieved by focusing only on the effective potential, which can be defined in our treatment, with a single-peak structure. Furthermore, adopting the geometric optics approximations, we consider the frequencies of eikonal QNMs and identify their correspondence with bound photon orbits in the SBH-disk model.

The rest of this paper is organized as follows. In sec.~\ref{sec.diskintro}, we briefly review the SBH-disk model proposed in Ref.~\cite{Kotlarik:2022spo}. In order to analyze the QNMs of the SBH-disk model, in sec.~\ref{sec.deformedBH} we consider a deformed Schwarzschild black hole, and demonstrate how to recast the master equation for scalar field perturbations in a Schr\"odinger-like form. This section is based on the results of Ref.~\cite{Chen:2022ynz}. The main results of our paper are presented in sec.~\ref{sec.BHdisk}, in which we show how the effective potentials of the master equation (sec.~\ref{subsec:potential}) and the QNM frequencies (sec.~\ref{subsec:qnm}) vary with respect to the parameters in the SBH-disk model. Then, in sec.~\ref{sec.eikonal}, we comment on the eikonal correspondence between QNMs and bound photon orbits in the SBH-disk model. Finally, we conclude in sec.~\ref{sec.conclusion}.

\section{The SBH-disk model}\label{sec.diskintro}

Due to the inherent non-linearity of Einstein equations, it is difficult to ``superpose" multiple sources in GR. However, in the static and axially symmetric case, the situation is much simpler. In fact, in Weyl cylindrical coordinates $(t, \rho, z, \varphi)$ Einstein equations outside of sources (i.e. in vacuum) are reduced to the Laplace equation and a line integration
\begin{align}
    \Delta \nu &= 0 \label{LaplaceEquation} \,, \\ 
    \lambda_{,\rho} &= \rho ( \nu_{,\rho}^2 - \nu_{,z}^2) \,, \qquad   \lambda_{,z} = 2\rho \nu_{,\rho} \nu_{,z} \,, \label{lambda}
\end{align}
where the $\nu(\rho, z)$ and $\lambda(\rho, z)$ are the only nontrivial components of the Weyl-type metric
\begin{equation}
	d s^2 = - e^{2\nu} d t^2 + \rho^2 e^{-2\nu} d\varphi^2 + e^{2\lambda - 2\nu} (d\rho^2 + d z^2) \,.
    \label{weylMetric}
\end{equation}
Thus any axially symmetric gravitational field with its potential $\nu$ known from Newton's theory has its GR counterpart. However, the potential $\nu$ does not tell the whole story. The presence of the second metric function $\lambda$ may significantly depart from the pure Newtonian picture. Moreover, while the Laplace equation \eqref{LaplaceEquation} is linear, and thus makes the superposition problem for the potential $\nu$ trivial, it is not the case for $\lambda$ as Eqs. \eqref{lambda} are quadratic in $\nu$.

Here, we wish to study the QNMs of a black hole that is surrounded by some matter in a physically appealing configuration. Namely, we take a recently derived solution \cite{Kotlarik:2022spo} describing a Schwarzschild black hole encircled by a thin disk (SBH-disk model). Such a structure is of clear astrophysical importance as disk-like sources often result from an accretion of matter onto a compact central body. While the total potential is a simple sum $\nu = \nu_\text{Schw} + \nu_\text{disk}$, for the second metric function we write $\lambda = \lambda_\text{Schw} + \lambda_\text{disk} + \lambda_\text{int}$, where $\lambda_\text{Schw}$ and $\lambda_\text{disk}$ denote contributions from the Schwarzschild black hole and the disk (thus each satisfying \eqref{lambda} with their corresponding $\nu_\text{Schw}$, or, $\nu_\text{disk}$ respectively). The non-linear ``interaction" part $\lambda_\text{int}$ satisfies
\begin{align}
    \lambda_{\text{int}, \rho} &= 2 \rho (\nu_{\text{Schw}, \rho} \nu_{\text{disk}, \rho} - \nu_{\text{Schw}, z} \nu_{\text{disk}, z}) \,, \\
	\lambda_{\text{int}, z} &= 2 \rho (\nu_{\text{Schw}, \rho} \nu_{\text{disk}, z} + \nu_{\text{Schw}, z} \nu_{\text{disk}, \rho}) \,.
\end{align}
Notice that when we treat the existence of the disk as a small perturbation of the black hole, i.e. $|\nu_\text{disk}| \ll |\nu_\text{Schw}|$, {and consider its contributions up to the first order, only the interaction part $\lambda_\text{int}$ is relevant because $\lambda_\text{disk}$ is of second order.

In Weyl coordinates, the Schwarzschild black hole is a singular rod of length $2M$ -- twice the black-hole mass $M$ -- placed symmetrically on the $z$ axis described by
\begin{align}
    \nu_\text{Schw} &= \frac{1}{2}\ln\left(\frac{R_++R_--2M}{R_++R_-+2M}\right)\,, \\
    \lambda_\text{Schw} &= \frac{1}{2}\ln\left[\frac{(R_++R_-)^2-4M^2}{4R_+R_-}\right] \,,
\end{align}
where
\begin{equation}
    R_\pm = \sqrt{\rho^2+(|z| \mp M)^2}\,.
\end{equation}
The disks considered in Ref.~\cite{Kotlarik:2022spo} are infinitesimally thin and spatially infinite (with a finite total mass) extending from the horizon. The disk density falls off quickly enough both at the horizon and at infinity -- see the schematic Fig. \ref{fig.superposition}. The Newtonian surface density profiles\footnote{The quantity $w(\rho)$ satisfies exactly the Poisson equation $\Delta \nu = 4 \pi w(\rho) \delta(z)$, where $\delta(z)$ is the delta distribution, so it is the precise counterpart of the Newtonian surface density.} read
\begin{equation}
	w^{(m, n)} = W^{(m, n)}\frac{b^{2m+1}\rho^{2n}}{2\pi(\rho^2 + b^2)^{m + n+ 3/2}} \,, \quad m,n \in \mathrm{N}_0
	\label{VLdensity}
\end{equation}
where $b$ is a parameter of the dimension of length and the normalization $W^{(m,n)}$ is chosen in such a way that the total mass of the disk $2\pi\int_0^\infty w^{(m,n)}(\rho) \rho \, d \rho = \mathcal{M}$. In particular,
\begin{equation}
    W^{(m,n)} =  (2m+1) \binom{m + n+ 1/2}{n} \mathcal{M}\,.
\end{equation}
The densities \eqref{VLdensity} have a single maximum located at $\rho_\text{max} = b \sqrt{\frac{2n}{3 + 2m}}$. Thus, increasing $b$ (or $n$) when $m,n$ (or $m,b$) are fixed means shifting the maximum further from the central region, as well as expanding the width of the peak. Whereas increasing $m$ when $n, b$ fixed corresponds to shifting the maximum towards the central region while shrinking the width of the peak. When keeping the total disk mass $\mathcal{M}$ constant, the maximum density decreases when increasing $b$ (or $n$) while it increases when increasing $m$.

\begin{figure} \includegraphics[width=\columnwidth]{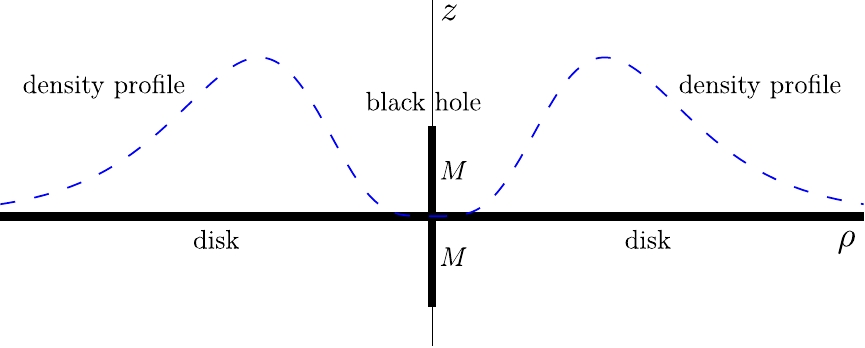}
    \caption{The schematic plot in Weyl coordinates of a Schwarzschild black hole (thick black vertical line) encircled by a thin disk \cite{Kotlarik:2022spo} (thick black horizontal line). The disk lies in the equatorial plane stretching from the horizon to infinity. The disk surface density profile \eqref{VLdensity} is indicated by the dashed blue line.} \label{fig.superposition}
\end{figure}

If we denote
\begin{equation}
	r_b^2 := \rho^2 + (|z| + b)^2 \;, \qquad |\cos\theta_b| := \frac{|z| + b}{r_b} \,,
 \label{rbthetab}
\end{equation} 
the potential is given by
\begin{equation}
	\nu^{(m, n)} = - W^{(m,n)}\sum_{j=0}^{m + n} \mathcal{Q}_j^{(m,n)} \frac{b^j}{r_b^{j+1}} P_j(|\cos\theta_b |) \,,
    \label{VLpotential}
\end{equation}
where $P_j$ are the Legendre polynomials and the coefficients
\[   
\mathcal{Q}_j^{(m,n)} = 
	\begin{cases}
		\sum_{k=0}^n (-1)^k \binom{n}{k} \frac{2^{j-k-m} (2m + 2k -j)!}{(m + k - j)!(2m + 2k + 1)!!} & \text{ if } j\leq m \\
		\sum_{k=j}^{m+n} (-1)^{k-m} \binom{n}{k-m} \frac{2^{j-k} (2k - j)!}{(k-j)!(2k + 1)!!} & \text{ if } j > m \,.
	\end{cases}
\]
The potential \eqref{VLpotential} was first obtained by Vogt \& Letelier \cite{Vogt:2009} by taking a specific superposition of the Kuzmin-Toomre family of discs \cite{Toomre:1963}.

The second metric function $\lambda_\text{disk}$ was also found explicitly (see \cite{Kotlarik:2022spo} Eq. (21)), but we will not repeat it here as we shall not need it. The interaction part $\lambda_\text{int}$ satisfies following recurrence relations
\begin{align}
    &\lambda^{(0,0)}_\text{int} = - \frac{\mathcal{M}}{r_b} \left( \frac{R_+}{b+M} - \frac{R_-}{b-M} \right) - \frac{ 2\mathcal{M} M}{b^2 - M^2}\,,\\
    &\lambda^{(0,n + 1)}_\text{int} =  \lambda^{(0,n)}_\text{int} + \frac{b}{2(n + 1)} \frac{\partial}{\partial b}\lambda^{(0,n)}_\text{int} \,, \\
    &\frac{(2m + 1)(2n + 3)}{2m + 2n + 3} \lambda_\text{int}^{(m+1, n)} \nonumber\\
    &=\lambda_\text{int}^{(m, n)} + \frac{4m(n+1)}{2m + 2n + 3} \lambda_\text{int}^{(m, n+1)} - b\frac{\partial}{\partial b} \lambda_\text{int}^{(m, n)}\,.
\end{align}
Thus the whole metric (both metric functions) of the SBH-disk model is known explicitly and in closed-form. From now on, to simplify the expression, the notation $(m,n)$ that indicates the explicit dependence of the disk functions on the indices $m$ and $n$ will be dropped. One should keep in mind that $\nu$ and $\lambda$ explicitly depend on $\mathcal{M}$, $b$, $m$, and $n$.

Two physical interpretations of these disks are possible: a) a single component ideal fluid with density $\sigma$ and azimuthal pressure $P$ (a set of solid rings with internal azimuthal stress), or, b) two equally counter-rotating pressureless dust streams with the densities $\sigma_\pm = \sigma/2$ following circular geodesics. Both characteristics follow from the metric
\begin{align}
    \sigma + P &= e^{\nu - \lambda} \frac{\nu_{,z}(z=0^+)}{2\pi} = e^{\nu - \lambda}  w(\rho) \,, \\ 
    P &= e^{\nu - \lambda}  \frac{\nu_{,z}(z=0^+)}{2\pi} \rho \nu_{,\rho} = e^{\nu - \lambda}  w(\rho) \rho \nu_{,\rho} \,,
    \label{physCharacteristics}
\end{align}
where $w(\rho)$ is the Newtonian surface density \eqref{VLdensity}. See Appendix \ref{app.diskProperties} for the derivation in more details.

Clearly $\sigma + P \geq 0$, so the strong energy condition is satisfied automatically for any disk. The dominant energy condition is generally satisfied everywhere (for a broad range of parameters) except close to the black-hole horizon, where $\sigma < P$. In fact, the accretion disks are usually assumed to end around the innermost stable circular orbit (ISCO). However, we argue that (i) our disk density drops to zero toward the horizon, so there is really no matter on the horizon itself, and, (ii) accretion disks around realistic black holes would indeed stretch toward the horizon, although the matter will infall there rather than orbiting on circular trajectories. Thus, in this sense, it is more realistic to model the gravitational field with some modest density going down to the horizon. By choosing appropriate parameters $(m,n)$ and $b$, the density can be made arbitrarily small below a chosen radius, e.g., the ISCO orbit.

For the double-stream interpretation, both energy conditions considered above require $\sigma_\pm \geq 0$, which also implies $P \geq 0$ for the single component interpretation. Finally, the energy conditions are satisfied for both interpretations if the speed of a particle on a circular geodesic in the equatorial plane 
\begin{equation}
    v^2 = \frac{P}{\sigma} = \frac{\rho \nu_{,\rho}}{1 - \rho \nu_{,\rho}}
\end{equation}
acquire timelike values $0 \leq |v| < 1$. 

While superposition can be carried out very straightforwardly in Weyl coordinates, it will be convenient to work in Schwarzschild coordinates ($t,r,\theta,\varphi$) from now on. The two sets of coordinates are related as follows\footnote{Note the difference between $r_b$ and $\theta_b$ with the subscript $b$ defined in \eqref{rbthetab} and the Schwarzschild coordinates $r, \theta$.}
\begin{equation}
    \rho = \sqrt{r (r-2M)} \sin\theta \,, \quad z = (r-M) \cos\theta \,,
\end{equation}
and the metric of the SBH-disk model in Schwarzschild coordinates then reads
\begin{align}
    ds^2=&-f(r)\textrm{e}^{2\nu_{\textrm{disk}}}dt^2+\textrm{e}^{2\lambda_{\textrm{ext}}-2\nu_{\textrm{disk}}}\frac{dr^2}{f(r)}\nonumber\\
    &+r^2\textrm{e}^{-2\nu_{\textrm{disk}}}\left(\textrm{e}^{2\lambda_{\textrm{ext}}}d\theta^2+\sin^2\theta d\varphi^2\right)\,,\label{superposedmetric}
\end{align}
where $\lambda_\textrm{ext}=\lambda_{\textrm{disk}}+\lambda_{\textrm{int}}$ and $f(r) \equiv \nu_\text{Schw} = 1 - 2M/r$ after the transformation into Schwarzschild coordinates.

\section{Deformed Schwarzschild black holes -- master equation}\label{sec.deformedBH}

The SBH-disk metric of Eq.~\eqref{superposedmetric} describes the spacetime of a Schwarzschild black hole encircled by a gravitating thin disk. The main goal of this work is to investigate the QNMs propagating in this superposed spacetime. However, due to the general $(r,\theta)$ dependence appearing in the metric functions through $\nu_\textrm{disk}$ and $\lambda_\textrm{ext}$, the radial and the latitudinal sectors of the wave equation are not separable. In order to proceed, we assume that the disk mass $\mathcal{M}$ is much smaller than the black hole mass $M$ and consider the contributions up to $O(\mathcal{M}/M)$. Besides having its astrophysical applicability, this assumption, as mentioned in the previous section, allows us to simplify the calculations by omitting $\lambda_\textrm{disk}$ term because it is of second order in $\mathcal{M}/M$. Then, we focus on the QNMs of scalar field perturbations. Adopting the projection method \cite{Cano:2020cao} to the master equation up to $O(\mathcal{M}/M)$, one can separate the radial component of the master equation from the latitudinal one. This has been shown explicitly in Ref.~\cite{Chen:2022ynz} for a very general class of deformed Schwarzschild spacetimes. In this section, we briefly review the results in Ref.~\cite{Chen:2022ynz}, based on which one can compute the scalar field QNMs of the SBH-disk model.

We consider a deformed Schwarzschild spacetime and assume that the spacetime remains static and axially symmetric in the presence of deformations. The nonzero metric components of the deformed spacetime can be expressed as \cite{Chen:2022ynz}
\begin{align}
g_{tt}(r,\theta)&=-f(r)\left(1+\epsilon A_j(r)|\cos^j\theta|\right)\,,\nonumber\\
g_{rr}(r,\theta)&=\frac{1}{f(r)}\left(1+\epsilon B_j(r)|\cos^j\theta|\right)\,,\nonumber\\
g_{\theta\theta}(r,\theta)&=r^2\left(1+\epsilon C_j(r)|\cos^j\theta|\right)\,,\nonumber\\ g_{\varphi\varphi}(r,\theta)&=r^2\sin^2\theta\left(1+\epsilon D_j(r)|\cos^j\theta|\right)\,,\label{deformedmetric}
\end{align}
where $\epsilon$ is a dimensionless parameter that quantifies the amount of deformations. In general, the spacetime deformations are functions of $r$ and $\theta$. In Eqs.~\eqref{deformedmetric}, we expand the latitudinal part of the deformation functions as a Taylor series in terms of $\cos\theta$. Each term in the series is weighted by a function of $r$, i.e., the functions $A_j(r)$, $B_j(r)$, $C_j(r)$, and $D_j(r)$ that appear in the expansion. The dummy index $j$ stands for summations running upward from $j=0$. The absolute value in each term in the expansion is to preserve the equatorial reflection symmetry, with the possibility of having a nonzero surface density at the equatorial plane. When the deformations are small, i.e., $|\epsilon|\ll1$, we can consider terms up to $O(\epsilon)$. As we will show later, the radial sector of the Klein-Gordon equation can then be separated from the latitudinal one, and it can be further recast into the Schr\"{o}dinger-like form. 

\subsection{Massless scalar field: Effective potential}

In this work, we will focus on the massless scalar field perturbations, whose QNMs are governed by the Klein-Gordon equation
\begin{equation}
\Box\psi=0\,.\label{KGeq}
\end{equation}
Indeed, the investigation of the ringdown phase in real gravitational wave emission has to be based on the computations of linearized gravitational equations rather than Eq.~\eqref{KGeq}. However, as the simplest scenario, the consideration of scalar field perturbations already allows us to address interesting issues, such as the (in)stability of the system, without suffering the computational complexity in linearized gravitational equations of deformed background spacetimes. In addition, according to the geometric optics approximations, the behaviors of scalar field QNMs should be able to capture those of the gravitational perturbations at least in the eikonal regimes, that is, when the multipole number $l$ is large. This will be discussed later in sec.~\ref{sec.eikonal}.  

For the master equation of scalar fields in the Schwarzschild spacetime, one can use the associated Legendre functions $P_l^{m_z}(x)$, where $x\equiv\cos\theta$ and $m_z$ is the azimuthal number, as the angular basis to separate the radial and latitudinal sectors of the wave equation. The radial equation is labeled by the multipole number $l$ and determines the evolution of the mode of $l$. The azimuthal number $m_z$ degenerates because of the spherical symmetry of the spacetime. In the presence of deformations of $O(\epsilon)$, there would appear off-diagonal terms that correspond to the modes with multipole numbers $l$ different from that of the zeroth-order one. These off-diagonal terms in the wave equations are $O(\epsilon)$. Therefore, by taking advantage of the orthogonality of $P_l^{m_z}$ among multipole numbers, one can project out the off-diagonal terms and focus only on the corrections on the zeroth-order equation. In the following, we will only show the main results of the calculations and refer the readers to sec. IV of Ref.~\cite{Chen:2022ynz} for more details.

Essentially, the projection method allows us to separate the radial and the latitudinal sectors of the wave equation. To further recast the radial equation into the Schr\"odinger-like form, we find it convenient to define the following coefficients:
\begin{align}
a_{l{m_z}}^j&=\frac{2{m_z}^2}{\mathcal{N}_{l{m_z}}}\int_0^1\frac{x^j\left(P_l^{m_z}\right)^2}{1-x^2}dx\,,\label{coeffa}\\
b_{l{m_z}}^j&=\frac{2}{\mathcal{N}_{l{m_z}}}\int_0^1x^j\left(P_l^{m_z}\right)^2dx\,,\label{coeffb}\\
c_{l{m_z}}^j&=\frac{2}{\mathcal{N}_{l{m_z}}}\int_0^1x^jP_l^{m_z}\left[\left(1-x^2\right)\partial_x^2-2x\partial_x\right]P_l^{m_z}dx\,,\label{coeffc}\\
d_{l{m_z}}^j&=\frac{2}{\mathcal{N}_{l{m_z}}}\int_0^1P_l^{m_z}\left(1-x^2\right)\left(\partial_xx^j\right)\left(\partial_x P_l^{m_z}\right)dx\,,\label{coeffd}
\end{align}
where the normalization constant $\mathcal{N}_{l{m_z}}\equiv 2(l+m_z)!/[(2l+1)(l-m_z)!]$ is determined by the orthogonality condition
\begin{equation}
\int_{-1}^1dx P_l^{m_z}(x)P_k^{m_z}(x)=\mathcal{N}_{l{m_z}}\delta_{lk}\,.
\end{equation}
Note that the coefficients given by Eqs.~\eqref{coeffa}-\eqref{coeffd} depend on $l$ and $m_z$, but they are invariant under $m_z\leftrightarrow-m_z$. 

After Fourier transformations, we denote the radial part of the Fourier modes of the scalar field as $\Psi_{l,m_z}(r)$. By using the projection method, the radial wave function is found to satisfy the following Schr\"{o}dinger-like equation \cite{Chen:2022ynz}
\begin{equation}
\partial_{r_*}^2\Psi_{l,m_z}(r)+\omega^2\Psi_{l,m_z}(r)=V_{\textrm{eff}}(r)\Psi_{l,m_z}(r)\,,\label{kgfoier}
\end{equation}
where $\omega$ is the mode frequency. The effective potential $V_{\textrm{eff}}(r)$ can be expressed as
\begin{widetext}
\begin{align}
V_{\textrm{eff}}(r)&=l(l+1)\frac{f(r)}{r^2}+\frac{f(r)}{r}\frac{df}{dr}\left[1+\epsilon b_{lm_z}^j\left(A_j(r)-B_j(r)\right)\right]\nonumber\\&+\epsilon\bigg\{\frac{f(r)}{r^2}\left[a_{lm_z}^j\left(A_j(r)-D_j(r)\right)-c_{lm_z}^j\left(A_j(r)-C_j(r)\right)-\frac{d_{lm_z}^j}{2}\left(A_j(r)+B_j(r)-C_j(r)+D_j(r)\right)\right]\nonumber\\&-\frac{b_{lm_z}^j}{4}\frac{d^2}{dr_*^2}\left[A_j(r)-B_j(r)\right]+\frac{1}{4r^2}\frac{d}{dr_*}\left[b_{lm_z}^jr^2\frac{d}{dr_*}\left(A_j(r)-B_j(r)+C_j(r)+D_j(r)\right)\right]\bigg\}\,,\label{veff1}
\end{align}
\end{widetext}
which explicitly contains the coefficients given by Eqs.~\eqref{coeffa}-\eqref{coeffd}. The tortoise radius $r_*$ is defined as follows
\begin{equation}
\frac{dr}{dr_*}=f(r)\left\{1+\frac{\epsilon}{2}b_{lm_z}^j\left[A_j(r)-B_j(r)\right]\right\}\,.\label{tortoiser}
\end{equation}
On the above equations \eqref{veff1} and \eqref{tortoiser}, the summations over $j$ are implicitly assumed. It can be seen that when $\epsilon=0$, the effective potential and the whole master equation reduce to those of the Schwarzschild spacetime. In this case, as we have mentioned, the azimuthal numbers $m_z$ degenerate, and Eq.~\eqref{kgfoier} is labeled only by $l$. However, in the presence of deformations, the spacetime is no longer spherically symmetric, hence the degeneracy among $m_z$ splits. Different values of $|m_z|$ in the range of $0\le|m_z|\le l$ give distinctive QNM frequencies.

\section{QNMs of SBH-disk model}\label{sec.BHdisk}

Having discussed the master equation of the scalar field perturbations in a general deformed Schwarzschild spacetime, we then consider the SBH-disk model whose metric is given by Eq.~\eqref{superposedmetric}. The SBH-disk model can also be treated as a deformed Schwarzschild spacetime whose deformations are caused by the thin disk. Typical mass $M$ of the astrophysical black hole is usually expected to dominate over the mass of the accretion disk $\mathcal{M}$. Therefore, it is natural to set $\epsilon=\mathcal{M}/M$ and consider terms up to $O(\mathcal{M}/M)$. As we have mentioned, in this linear approximation, we have $\lambda_\textrm{ext}\approx\lambda_{\textrm{int}}$ because $\lambda_\textrm{disk}$ is quadratic in $\epsilon$. The metric components of the SBH-disk model can then be approximated as
\begin{align}
g_{tt}(r,\theta)&\approx-f(r)\left(1+2\nu_{\textrm{disk}}\right)\,,\nonumber\\ g_{rr}(r,\theta)&\approx \frac{1}{f(r)}\left(1+2\lambda_{\textrm{int}}-2\nu_{\textrm{disk}}\right)\,,\nonumber\\
g_{\theta\theta}(r,\theta)&\approx r^2\left(1+2\lambda_{\textrm{int}}-2\nu_{\textrm{disk}}\right)\,,\nonumber\\ g_{\varphi\varphi}(r,\theta)&\approx r^2\sin^2\theta\left(1-2\nu_{\textrm{disk}}\right)\,.\label{bhdiskmetric}
\end{align}
The approximated metric \eqref{bhdiskmetric} belongs to the class of deformed Schwarzschild metrics of Eq.~\eqref{deformedmetric}, as will be shown more explicitly below. 

\subsection{Effective potential}\label{subsec:potential}

The identification between the metrics \eqref{deformedmetric} and \eqref{bhdiskmetric} is made by first expanding $\nu_\textrm{disk}$ and $\lambda_\textrm{int}$ in terms of $|x|$ as follows:
\begin{equation}
\nu_{\textrm{disk}}=\epsilon \mathcal{V}_j(r)|x^j|\,, \quad \lambda_{\textrm{int}}=\epsilon \mathcal{L}_j(r)|x^j|\,,
\end{equation}
where $\mathcal{V}_j(r)$ and $\mathcal{L}_j(r)$ depend on $m$, $n$, and $b$, but are independent of $\epsilon$. Again, the summations over $j$ are implicitly imposed as before. One then identifies the weighting functions in Eq.~\eqref{deformedmetric} as follows
\begin{align}
A_k(r)&=-D_k(r)=2\mathcal{V}_k(r)\,,\nonumber\\
B_k(r)&=C_k(r)=2\mathcal{L}_k(r)-2\mathcal{V}_k(r)\,,\nonumber
\end{align}
for all $k$. With these mappings, one sees that the approximated metric \eqref{bhdiskmetric} does belong to the class of metrics \eqref{deformedmetric}. As a result, the effective potential \eqref{veff1} can be written as
\begin{widetext}
\begin{align}
V_{\textrm{eff}}(r)=l(l+1)\frac{f(r)}{r^2}&+\frac{f(r)}{r}\frac{df}{dr}\left[1+\epsilon b_{lm_z}^j\left(4\mathcal{V}_j(r)-2\mathcal{L}_j(r)\right)\right]\nonumber\\&+\epsilon\left\{\frac{f}{r^2}\left[4a_{lm_z}^j\mathcal{V}_j(r)-c_{lm_z}^j\left(4\mathcal{V}_j(r)-2\mathcal{L}_j(r)\right)\right]-\frac{b_{lm_z}^j}{2}\frac{d^2}{dr_*^2}\left[2\mathcal{V}_j(r)-\mathcal{L}_j(r)\right]\right\}\,,\label{veff2}
\end{align}
\end{widetext}
and the definition of the tortoise radius $r_*$, which is given by Eq.~\eqref{tortoiser}, becomes
\begin{equation}
\frac{dr}{dr_*}=f(r)\left\{1+\epsilon b_{lm_z}^j\left[2\mathcal{V}_j(r)-\mathcal{L}_j(r)\right]\right\}\,.\label{tortoiser2}
\end{equation}
Note that when $\epsilon=0$, the spacetime recovers a pure Schwarzschild one and the effective potential is given by
\begin{equation}
V_{\textrm{eff}}^{\textrm{Sch}}(r)\equiv l(l+1)\frac{f(r)}{r^2}+\frac{f(r)}{r}\frac{df}{dr}\,.
\end{equation}
With the master equation \eqref{kgfoier} and the effective potential \eqref{veff2}, we can calculate the QNM frequencies of the scalar field perturbations of the SBH-disk model.

\begin{figure}[t]
    \centering  \includegraphics[scale=0.32]{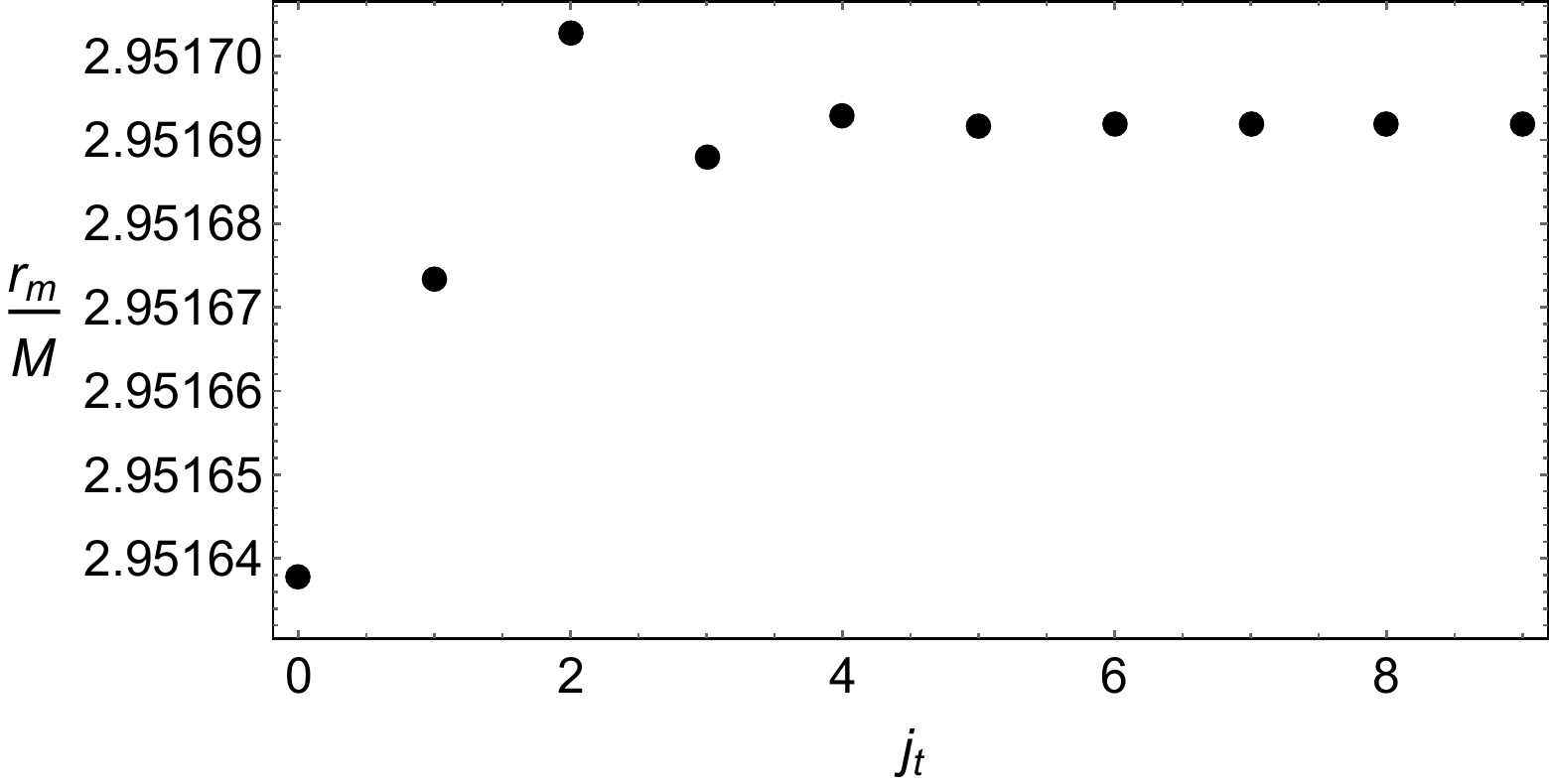}
    \caption{The radius of potential peak $r_m$ calculated with various $j_t$ at which the summation is truncated. In this figure we choose $m=0$, $n=1$, $b=10M$, $l=|m_z|=2$, $\mathcal{M}=0.02M$. The results converge very well already at $j_t=4$.}     \label{fig.pi}
\end{figure}

We first check that the effective potential, which is ideally defined as an infinite sum in $j$, has a sufficiently fast rate of convergence when increasing the summation order $j$. In Fig.~\ref{fig.pi}, we consider the effective potential $V_\textrm{eff}(r)$ of the SBH-disk model with $m=0$, $n=1$, $b=10M$, $l=|m_z|=2$, and $\mathcal{M}=0.02M$, then calculate the radius of its peak $r_m$ with various truncation values of $j$, which we defined as $j_t$. We find that when $j_t\ge4$, the results of $r_m$ already converge very well. Therefore, in the rest of this paper, the effective potential of the SBH-disk model will be calculated with the summation truncated at $j_t=4$.

\begin{figure}
    \centering  \includegraphics[scale=0.32]{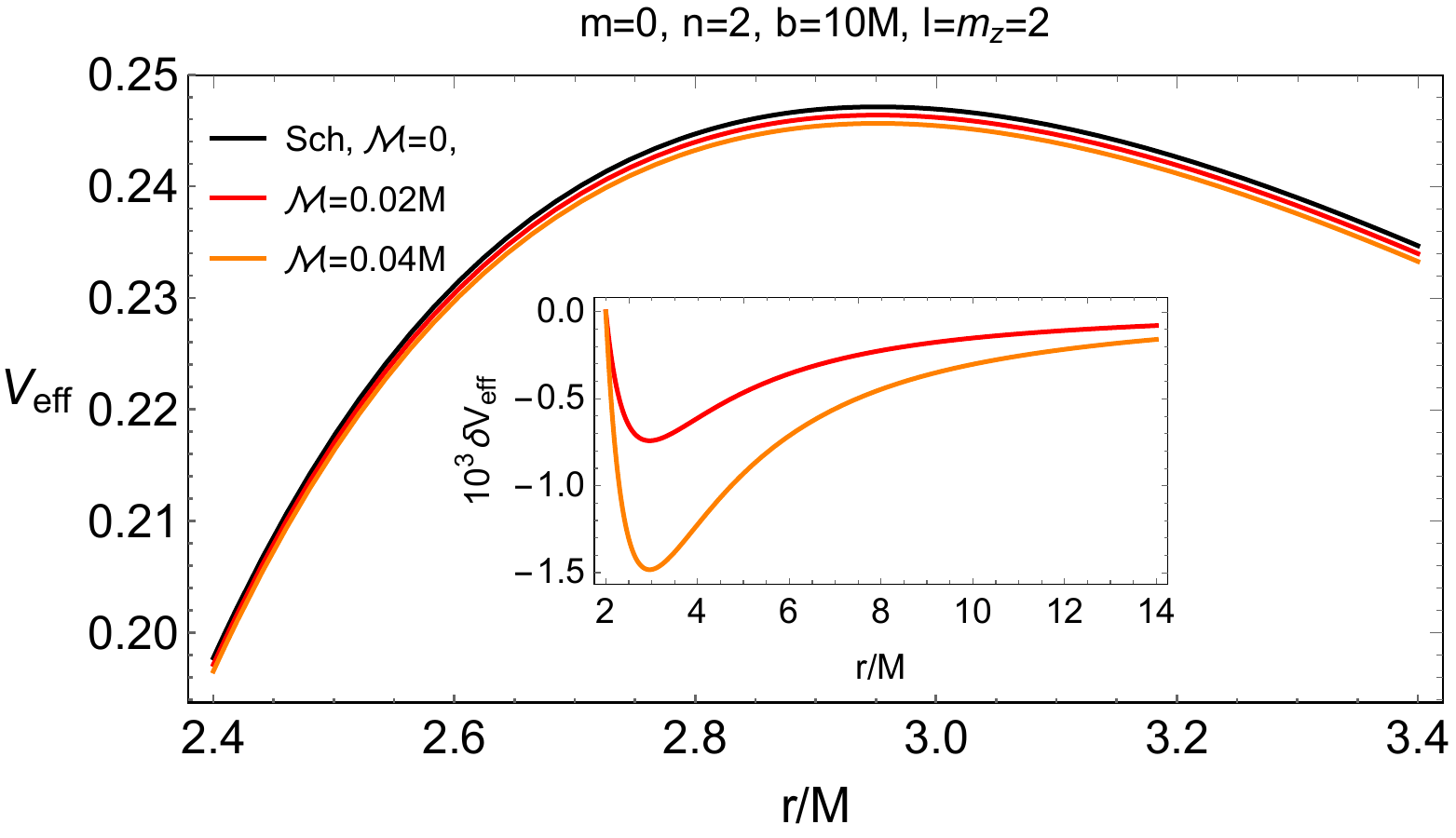}
    \caption{The effective potentials $V_{\textrm{eff}}(r)$ of the SBH-disk model are shown with various $\mathcal{M}/M$. The presence of the disk flattens the effective potential.}     \label{fig.pM}
\end{figure}

In Fig.~\ref{fig.pM}, we set $m=0$, $n=2$, $b=10M$, and $l=|m_z|=2$. The effective potentials of SBH-disk models are shown with respect to different values of the disk mass $\mathcal{M}$. The black curve corresponds to the pure Schwarzschild black hole, i.e., $\mathcal{M}=0$, whose effective potential is given by $V_\textrm{eff}^\textrm{Sch}(r)$. The inset shows the deviation of the effective potentials in the presence of the disk with respect to the pure Schwarzschild one ($\delta V_{\textrm{eff}}\equiv V_{\textrm{eff}}-V_{\textrm{eff}}^{\textrm{Sch}}$). One can see that the effective potential is flattened in the presence of the disk. This is consistent with the findings in Ref.~\cite{Kotlarik:2022spo} that the disk provides additional gravitational attractions and makes the horizon, as well as the effective potential as a whole, more flattened. Also, from the inset, one finds that the effective potential reduces to $V_\textrm{eff}^\textrm{Sch}(r)$ both near the horizon and at the spatial infinity. This is also expected as the surface density of the disk drops to zero there, as one can see in Fig.~\ref{fig.superposition}. In fact, the inset of Fig.~\ref{fig.pM} also indicates that the effective potential in the presence of the disk acquires the largest deviation from the pure Schwarzschild one near the peak $r_m$.

\begin{figure}
    \centering  \includegraphics[scale=0.32]{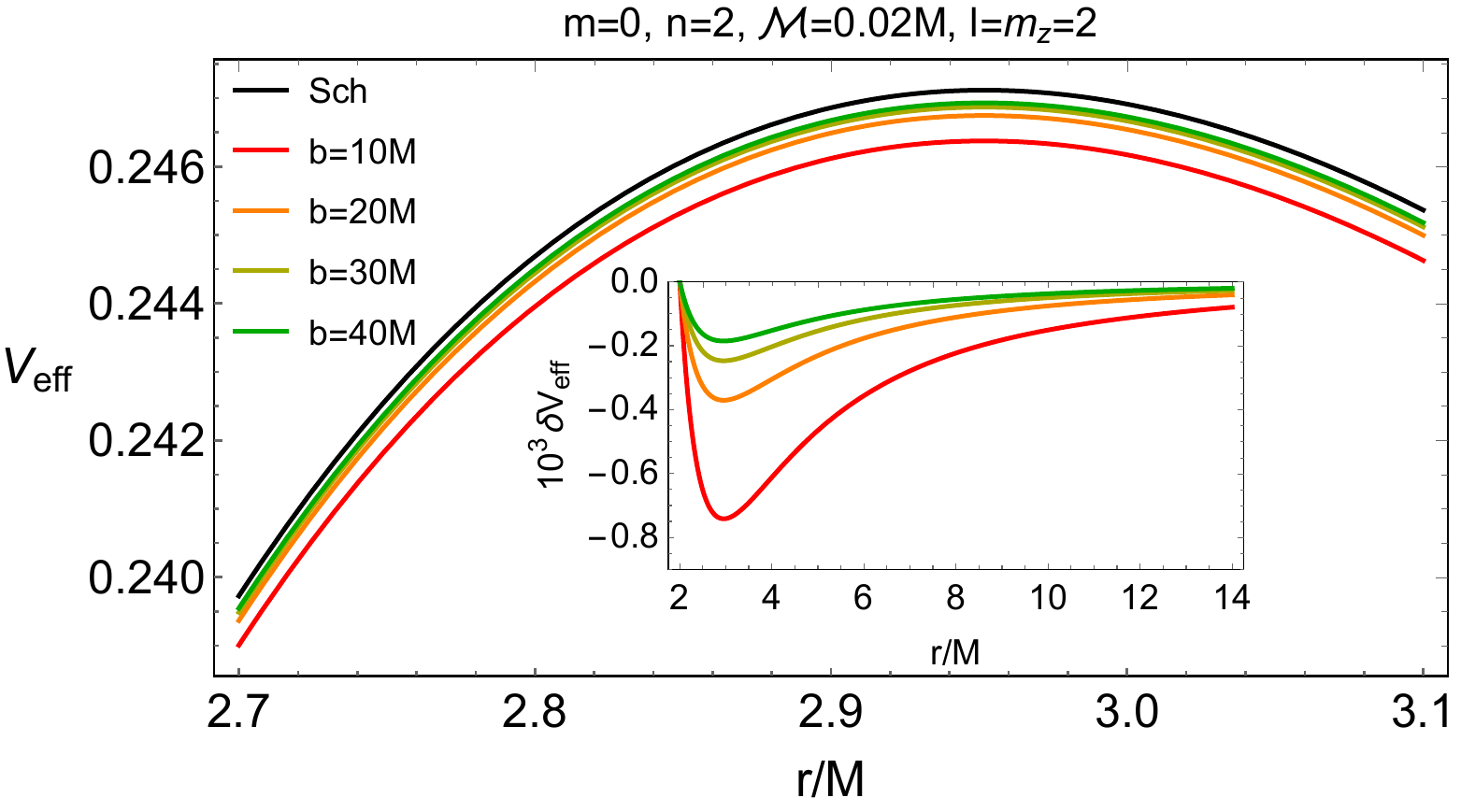}
    \caption{The effective potentials $V_{\textrm{eff}}(r)$ of the SBH-disk model are shown with various $b/M$. When $b/M$ increases, the effective potential converges to that of the pure Schwarzschild black hole.}     \label{fig.pb}
\end{figure}

Then, we explore the shape of the effective potential within the parameter space of the disk model itself, i.e., $m$, $n$, and $b$. In Figs.~\ref{fig.pb}, \ref{fig.pn}, and \ref{fig.pmvalue2}, we focus on how the effective potentials vary with respect to the changes of $b$, $n$, and $m$, respectively. We find that the effective potentials gradually reduce to $V_{\textrm{eff}}^{\textrm{Sch}}$ when increasing $b$ or $n$. On the other hand, increasing the index $m$ would further flatten the effective potential, as can be seen from Fig.~\ref{fig.pmvalue2}. As we have mentioned in sec.~\ref{sec.diskintro}, when keeping $\mathcal{M}/M$ constant and increasing either $n$, $b$, or $1/m$, the density peak of the disk would get lower and move further away from the black hole. Therefore, the net effects due to the disk become weaker. It is also worth remarking that in the presence of the disk, $|\delta V_{\textrm{eff}}|$ seems to always get its largest value near the potential peak $r_m$.

\begin{figure}
    \centering  \includegraphics[scale=0.32]{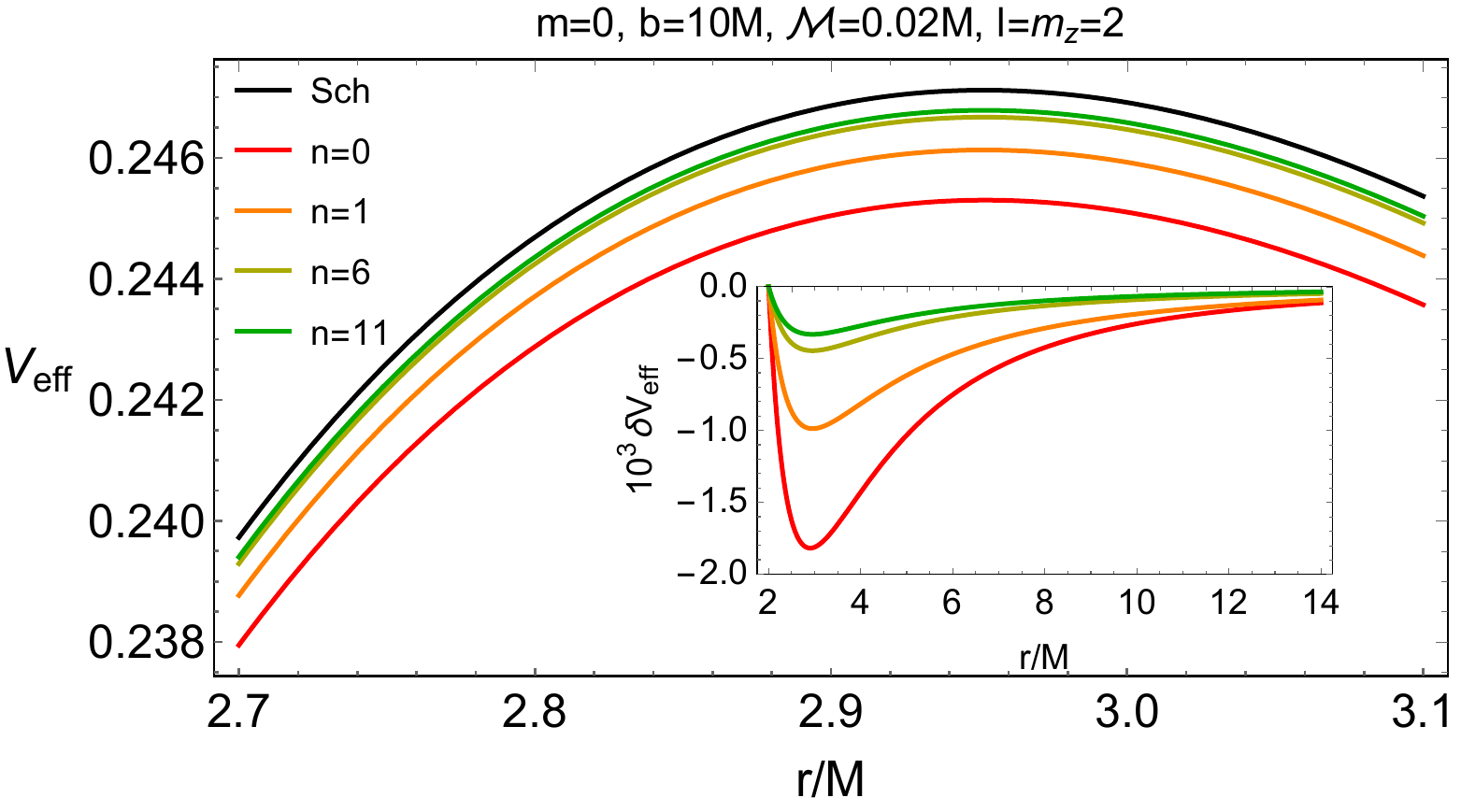}
    \caption{The effective potentials $V_{\textrm{eff}}(r)$ of the SBH-disk model are shown with different $n$. As $n$ increases, the effective potential slowly converges to the Schwarzschild one.}     \label{fig.pn}
\end{figure}

\begin{figure}
    \centering  \includegraphics[scale=0.32]{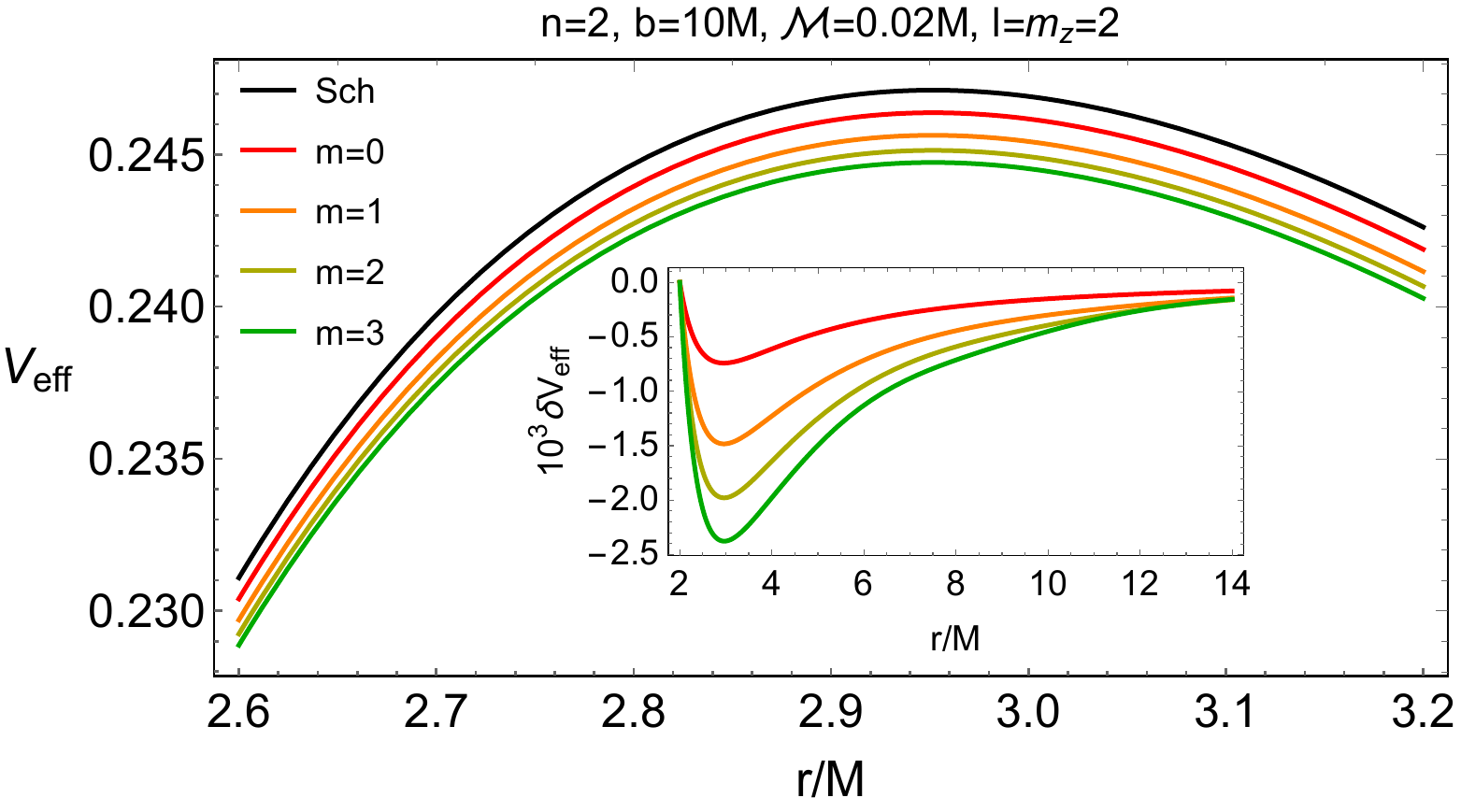}
    \caption{The effective potentials $V_{\textrm{eff}}(r)$ of the SBH-disk model are shown with different $m$. Increasing $m$ makes the peak of the surface density of the disk higher, narrower, and closer to the black hole, thus altering the effective potential more significantly.}     \label{fig.pmvalue2}
\end{figure}

\subsection{Scalar field QNMs}\label{subsec:qnm}

The QNM frequencies of the scalar field perturbations in the SBH-disk model can be calculated by solving Eq.~\eqref{kgfoier} with the effective potential \eqref{veff2} after imposing proper boundary conditions. Typical boundary conditions for black hole QNMs require that there are purely outgoing waves at spatial infinity and purely ingoing waves at the event horizon. The system can be treated as a wave-scattering problem through the peak of the effective potential. The whole system is dissipative because of the boundary conditions. Therefore, the QNM frequencies in general would acquire an imaginary part that quantifies the decay of the modes.

In this section, we focus on the cases where the effective potential retains its single-peak structure. This can be easily achieved when only orders of $O(\epsilon)$ are considered. The single-peak structure of the effective potential allows us to calculate the QNM frequencies using the third-order Wentze-Kramers-Brillouin (WKB) method \cite{Schutz:1985km,Iyer:1986np}{\footnote{The WKB method for calculating black hole QNMs has been extended to higher orders \cite{Konoplya:2003ii,Matyjasek:2017psv,Matyjasek:2019eeu,Hatsuda:2019eoj}. We refer the readers to Ref.~\cite{Konoplya:2019hlu} for the review of the method and, in particular, its range of applicability.}}. We also make use of the asymptotic iteration method (AIM) \cite{Cho:2009cj,Cho:2011sf} to check the consistency of the results. In this section, we shall focus on the fundamental modes with $l=|m_z|$ because the fundamental modes have the longest decay time and hence are more astrophysically relevant. In addition, our numerical results suggest that changing $|m_z|$ only shifts the frequencies very weakly as compared to the frequency shifts generated by other model parameters.

The complex planes of QNM frequencies within some parameter space are shown in Fig.~\ref{fig.qnmcal}. In each panel, the three branches correspond to the complex QNM frequencies with multipole numbers $l=2$, $l=4$, and $l=6$, from left to right, respectively. For each branch in the top panel, we fix the set of parameters $\{m,n,b\}$ as that in Fig.~\ref{fig.pM} and use the third-order WKB method to calculate the QNM frequencies with respect to the disk mass, which is chosen to be $\mathcal{M}=0$, $0.02M$, $0.04M$, $0.06M$, $0.08M$, and $0.1M$ (black points from top to bottom). The green (magenta) points are the results calculated using AIM, with disk mass $\mathcal{M}=0$ ($\mathcal{M}=0.1M$). In this case, the topmost points in each branch correspond to the QNM frequencies of a pure Schwarzschild black hole. In the bottom panel of Fig.~\ref{fig.qnmcal}, we fix the set of parameters $\{m,n,\mathcal{M}\}$ as that in Fig.~\ref{fig.pb}, and vary only $b=25M$, $20M$, $15M$, $10M$, and $5M$ (black points from top to bottom) for each branch. The green (magenta) points are the results calculated using AIM, with $b=5M$ ($b=25M$).

\begin{figure}
    \centering  \includegraphics[scale=0.32]{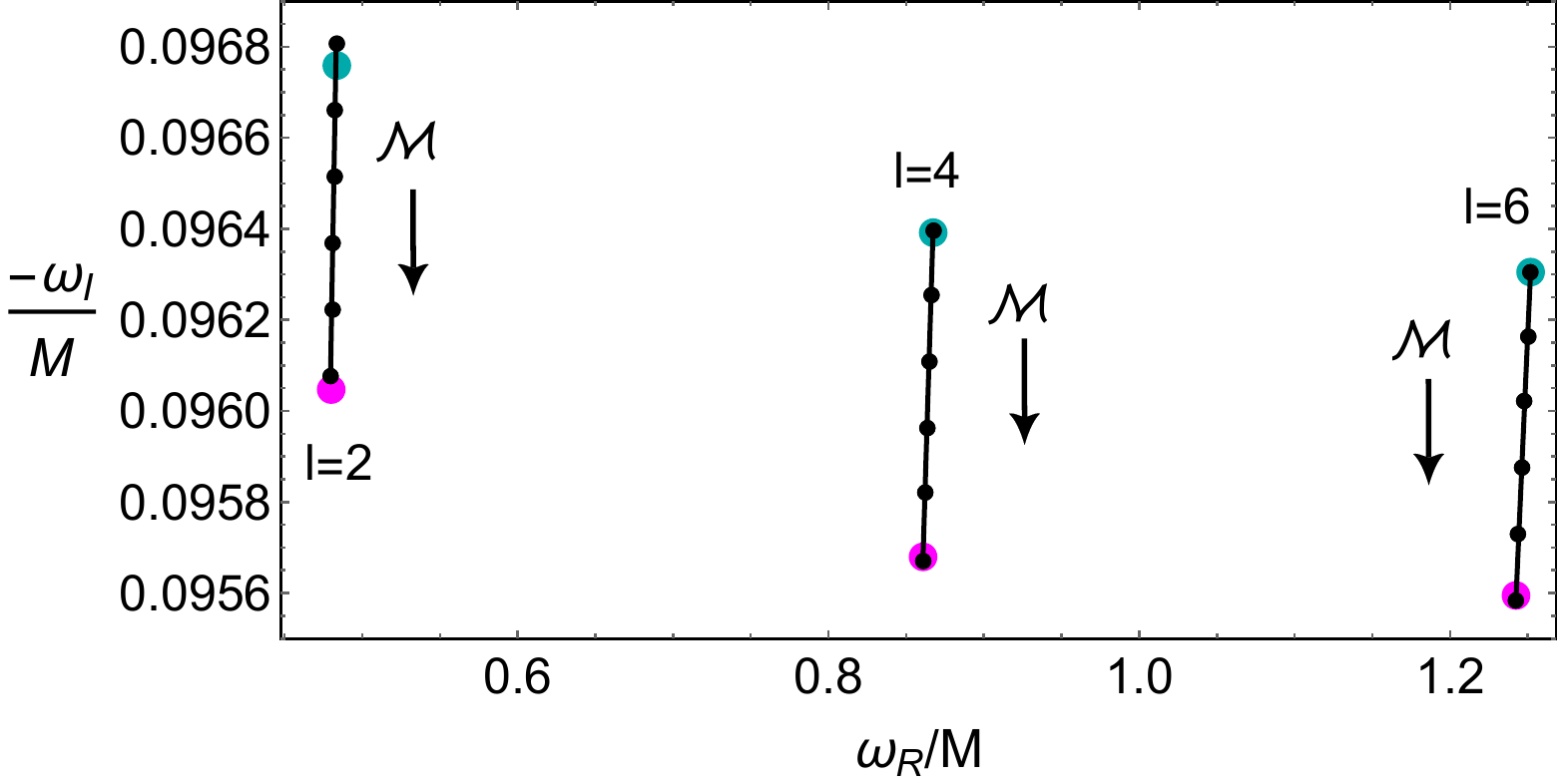}
    \includegraphics[scale=0.32]{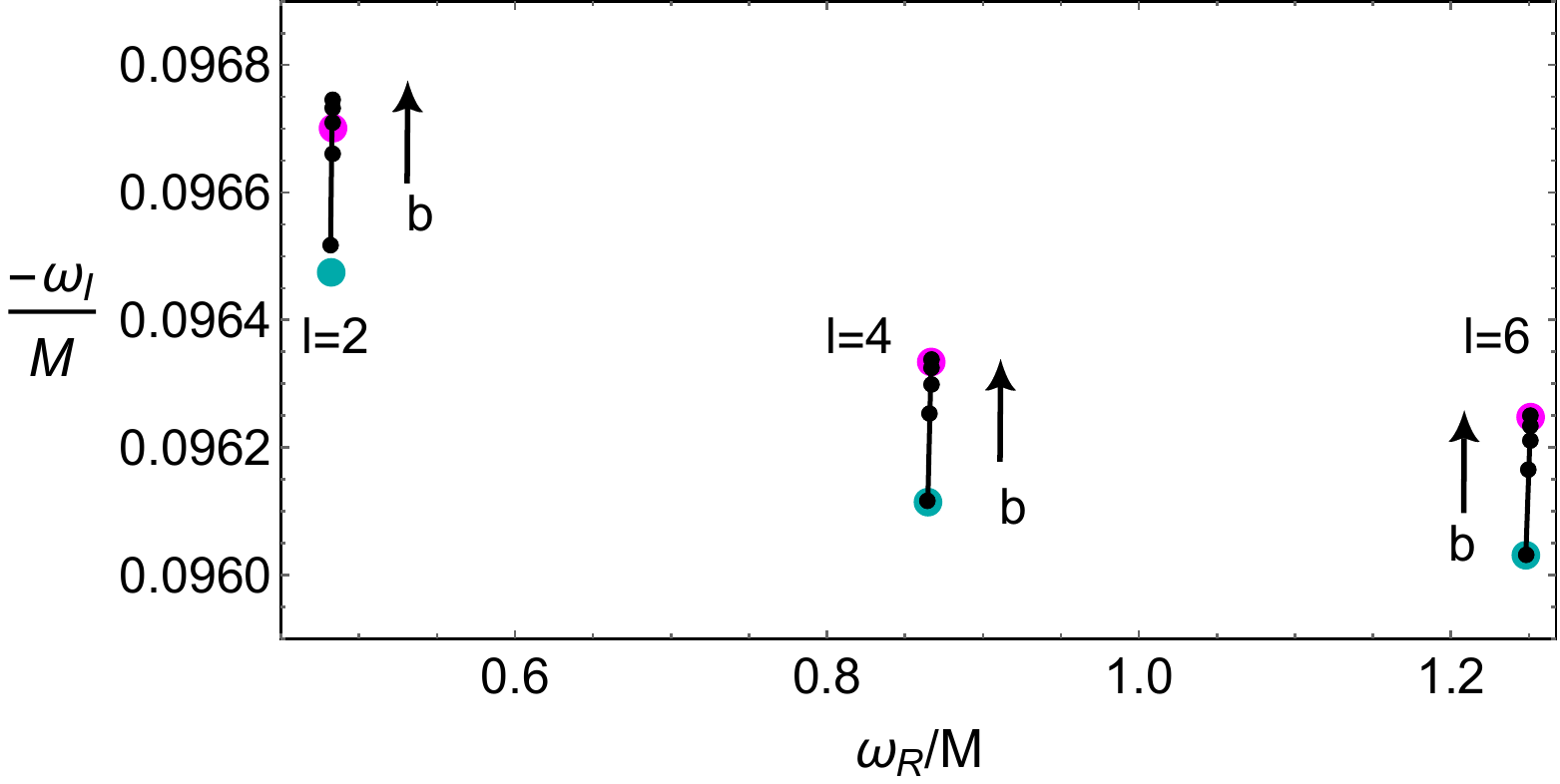}
    \caption{The QNM frequencies of the SBH-disk model with different choices of the parameters. The frequencies are calculated using the third-order WKB method (black points) and AIM (colored points). (Top) $\mathcal{M}$ varies from $0$ to $0.1M$ (from top to bottom in each branch) with other parameters $\{m,n,b\}$ fixed as those in Fig.~\ref{fig.pM}. The green and magenta points correspond to $\mathcal{M}/M=0$ and $0.1$, respectively. (Bottom) $b/M$ varies from $5$ to $25$ (from bottom to top in each branch) with other parameters $\{m,n,\mathcal{M}\}$ fixed as those in Fig.~\ref{fig.pb}. The green and magenta points correspond to $b/M=5$ and $25$, respectively. The arrows indicate the direction along which the referred parameters increase.}     \label{fig.qnmcal}
\end{figure}

From Fig.~\ref{fig.qnmcal}, one first sees that the WKB method and AIM give quite consistent results, particularly in the regime of large $l$ where the WKB method is expected to be accurate. Second, increasing the disk mass $\mathcal{M}$ would reduce the values of $\omega_R$ and $|\omega_I|$ {\footnote{The change of $\omega_R$ in each branch may not be easily seen in Fig.~\ref{fig.qnmcal}. See Fig.~\ref{fig.qnmdeg} for more details.}. In addition, given a non-zero $\mathcal{M}$, the pure Schwarzschild results can be recovered when $b\rightarrow\infty$. Reducing the value of $b$ decreases the values of $\omega_R$ and $|\omega_I|$, as compared with the pure Schwarzschild case. This is consistent with our previous finding that when $b$ increases, the effective potential gradually reduces to $V_\textrm{eff}^\textrm{Sch}$.

In fact, after a careful examination of the parameter space, we find that the presence of a thin disk would always reduce the values of $\omega_R$ and $|\omega_I|$, as long as $\mathcal{M}/M$ stays reasonably small and the index $m$ remains $O(1)$. In such cases, the validity of the first-order approximation used to derive the effective potential \eqref{veff2} is ensured. Moreover, the effective potential has a single-peak structure, whose shape monotonically deviates from the pure Schwarzschild one. In Fig.~\ref{fig.qnmdeg}, we focus on $l=2$ and investigate the QNM frequencies in the parameter space $\{m,n,b,\mathcal{M}\}$. The solid curve shows the results of fixing $m=0$, $n=2$, $b=10M$, and varying $\mathcal{M}/M$ from $0$ to $0.1$. The cross indicates the pure Schwarzschild frequency $\mathcal{M}=0$. The colored points and open circles show the results of fixing $\mathcal{M}/M=0.02$ and other parameters except for those indicated in the legend. From Fig.~\ref{fig.qnmdeg}, we find that the larger the parameters $b$ (red circle) or $n$ (green circle) are, the closer the QNM frequencies are to the Schwarzschild one. On the other hand, increasing $m$ would reduce $\omega_R$ and $|\omega_I|$.  
 
\begin{figure}
    \centering  \includegraphics[scale=0.32]{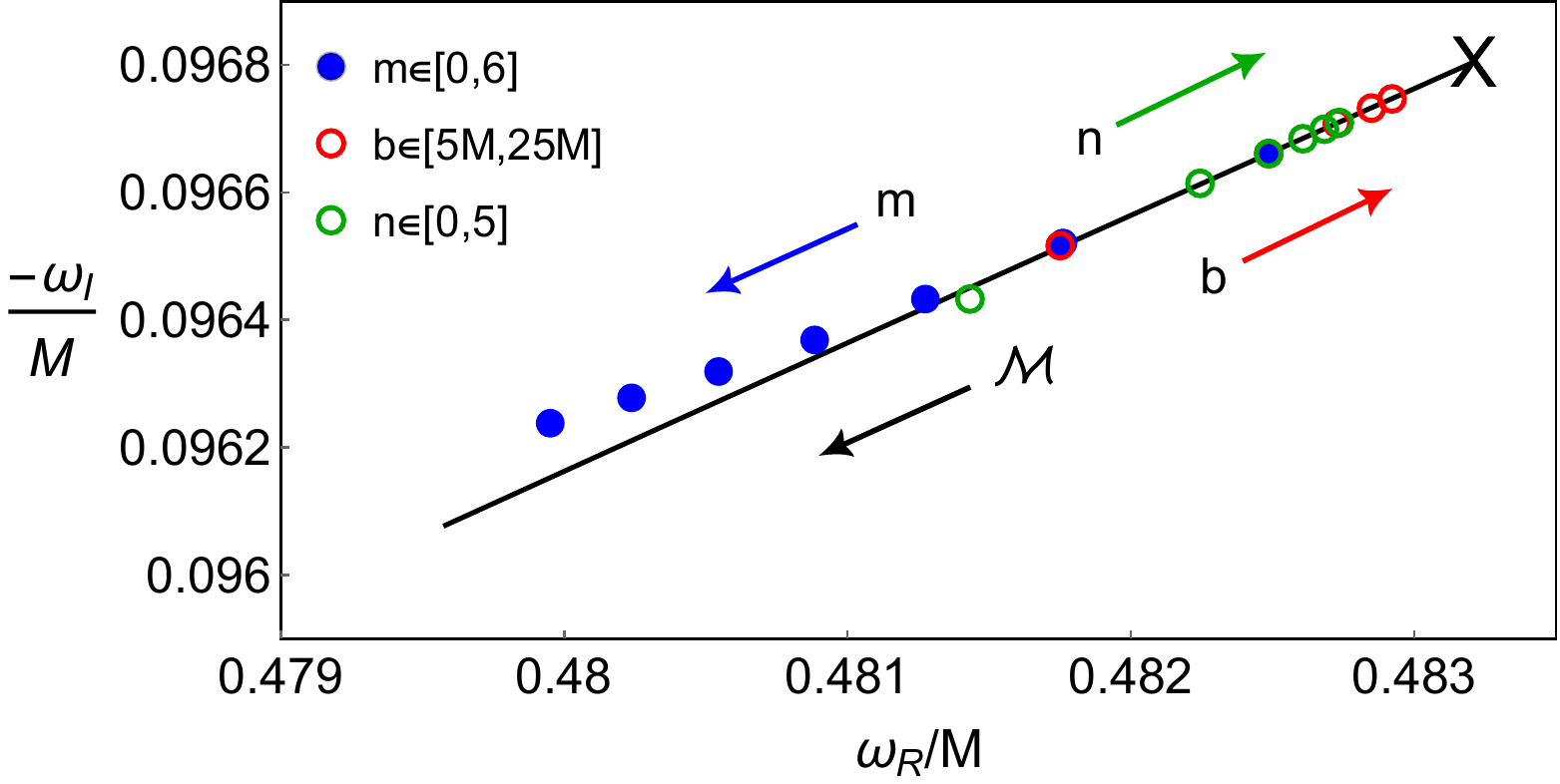}
    \caption{The $l=2$ QNM frequencies of the SBH-disk model. The solid curve shows the results of fixing $\{m,n,b\}$ as in Fig.~\ref{fig.pM}, while varying $\mathcal{M}/M$ from $0$ to $0.1$ (right to left). The cross corresponds to the pure Schwarzschild results $(\mathcal{M}=0)$. The colored points show the results of fixing $\mathcal{M}/M=0.02$ and other parameters except for the one indicated in the legend. The arrows indicate the direction along which the referred parameters change.}     \label{fig.qnmdeg}
\end{figure}

Another important observation from Fig.~\ref{fig.qnmdeg} is that almost all the colored points are nicely lined along the black curve. Although the SBH-disk model has a large parameter space $\{m,n,b,\mathcal{M},l,m_z\}$, there seems to be a universal relation that the QNM frequencies of the model have to obey. We also consider other multipole numbers $l$ and a universal relation seems to exist among the modes, as can be seen in Fig.~\ref{fig.qnmuni}. A similar trend of QNM shifts also appears in the model in which the black hole spacetime is superposed with a spherically symmetric matter distribution \cite{Cardoso:2021wlq,Konoplya:2021ube}, indicating that the relation may be really universal in the sense that it is insensitive to the matter configuration in the distribution. In general, the universal relation inevitably implies a strong degeneracy among intrinsic disk parameters. However, the relation could be helpful to distinguish the disk effects from those contributed by other putative external parameters not belonging to the disk model. For example, if the black hole QNM frequencies are found to be away from this universal relation, e.g, they are not lined along the black curve in Fig.~\ref{fig.qnmuni}, such a frequency shift must be induced by effects other than the disk contributions. In fact, several quantum-corrected black hole models predict larger values of $\omega_R$ \cite{Liu:2012ee,Fernando:2012yw,Flachi:2012nv,Bouhmadi-Lopez:2020oia,Daghigh:2020fmw,Jafarzade:2021umv,del-Corral:2022kbk}, hence the QNMs of the models would not be lined on the black curve. The quantum parameters in these models are thus robustly disentangled from the disk effects, hence enhancing the possibility of testing these quantum-corrected black hole models through black hole spectroscopy.  

Having said that, if there does exist a universal relation, one should still be careful with the range of its validity. In particular, from Fig.~\ref{fig.qnmuni}, the relation seems not valid anymore when one keeps increasing $m$ or $\mathcal{M}$. Indeed, disk models with sufficiently large $m$ and $\mathcal{M}$ could acquire a very dense and narrow peak in the density profile outside the black hole, resembling a flattened torus or ring rather than a disk. This extreme density profile could largely alter the shape of the effective potential, including the possibility of generating extra peaks outside the original one (see Fig.~\ref{fig.multipolepeak} for an example). If this happens, pseudospectral instability may be triggered, which would totally destroy the QNM spectrum \cite{Cheung:2021bol} and the universal relation would not be valid anymore. In fact, when the second peak appears in the effective potential, gravitational echoes following the main sinusoidal-decaying phase may appear in the time domain signals. These echoes correspond to the long-lived modes which are trapped between the potential barriers before they slowly leak through the outer one. However, we would like to emphasize that the possibility of having multiple peaks in the effective potential has to be treated with great care. This is because increasing the disk mass $\mathcal{M}$ would, at some point, violate the validity of the first-order approximations from which we derive the effective potential \eqref{veff2}. In addition, although all energy conditions discussed at the end of sec. \ref{sec.diskintro} hold for the disk parameters considered in Fig. \ref{fig.multipolepeak}, for high $m$ the density peak is located around the Schwarzschild value of the ISCO radius. Thus, for the most part, those disks would not be stable. Therefore, the double-peak structure in the effective potential demonstrated in Fig.~\ref{fig.multipolepeak} may have issues regarding its theoretical and physical viability. Therefore, we will not discuss it more in the present paper.     

\begin{figure}
    \centering  \includegraphics[scale=0.32]{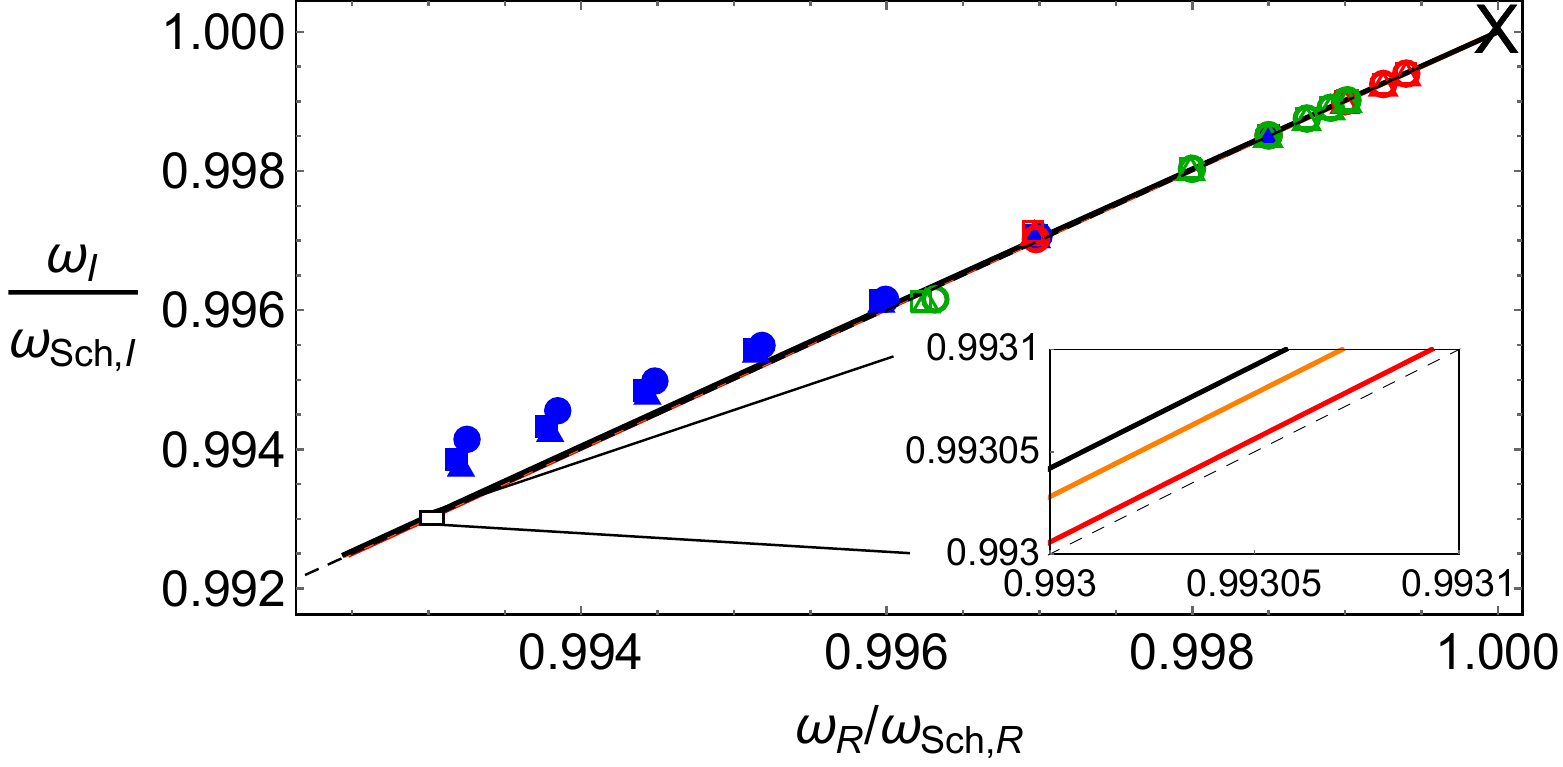}
    \caption{The ratio of the QNM frequencies of the SBH-disk model with various parameter choices, with respect to the Schwarzschild QNM frequencies $\omega_\textrm{Sch}$. The black ($l=2$), orange ($l=4$), and red ($l=6$) continuous curves show the results of varying $\mathcal{M}/M$ from $0$ to $0.1$. The cross located at the coordinate $(1,1)$ corresponds to the pure Schwarzschild results $(\mathcal{M}=0)$. The circular, rectangular, and triangular points correspond to $l=2$, $l=4$, and $l=6$, respectively. The parameter sets chosen for the points are referred to by the colors as those in Fig.~\ref{fig.qnmdeg}. The thin dashed line with a slope equal to one is shown for reference.}     \label{fig.qnmuni}
\end{figure}

\begin{figure}
    \centering  \includegraphics[scale=0.32]{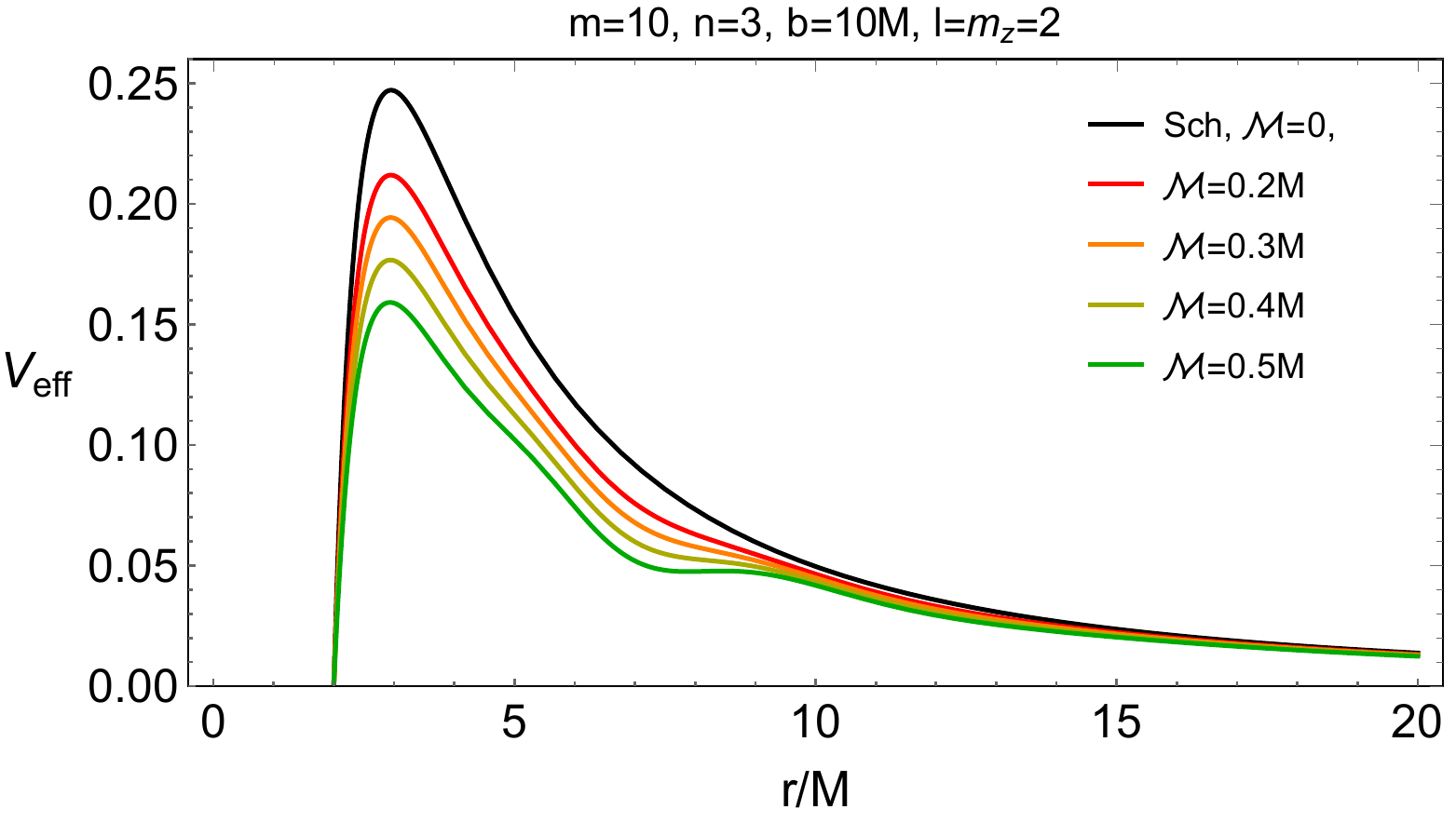}
    \caption{The effective potential $V_{\textrm{eff}}(r)$ of the SBH-disk model with $b=10M$, $n=3$, $m=10$, $l=|m_z|=2$. In some cases, the disk density is huge enough to induce multiple peaks in the effective potential (green).}     \label{fig.multipolepeak}
\end{figure}

\section{QNMs in Eikonal limits}\label{sec.eikonal}

Consider a test field propagating in curved spacetimes. In the geometric optics approximation, the wavelength of the field is assumed to be much smaller than any other length scale in the system. In the leading order of this approximation, sometimes also called eikonal approximation, the equations of motion of the propagating field share the same form as those of the freely moving photons. When adopting the approximation to black hole spacetimes, it is well-known that the eikonal black hole QNMs have some properties that can be directly linked to the photon orbits in such a spacetime. More explicitly, one can identify the so-called eikonal correspondence between the eikonal QNMs and the bound photon orbits around the black hole.

In a static and spherically symmetric black hole spacetime, the QNMs are determined up to their multipole number $l$ because the azimuthal numbers $m_z$ degenerate. As for the bound photon orbits, it turns out that all the bound photon orbits in this case are circular orbits and have a single radius, called the photon sphere \cite{Claudel:2000yi}. In this simple spacetime configuration, the eikonal correspondence can be identified straightforwardly through the fact that the peak of the effective potential of QNMs of $l\gg1$ is precisely at the photon sphere. Based on this identification, the real and the imaginary parts of the large-$l$ QNMs would correspond to the orbital frequency and the Lyapunov exponent of photons on the photon sphere, respectively \cite{Cardoso:2008bp}. The eikonal correspondence can be extended to rotating black hole spacetimes \cite{Yang:2012he,Li:2021zct}, black holes with multiple photon spheres \cite{Guo:2021enm}, and even deformed black hole spacetimes \cite{Chen:2022ynz}. The possibility of testing eikonal correspondence through black hole observations has been proposed in Ref.~\cite{Chen:2022nlw}.

When the black hole is slightly deformed, as the SBH-disk model considered in this paper, the QNM equations depend on both $l$ and $m_z$. Therefore, the identification of the eikonal correspondence has to be carried out with care. In fact, for the SBH-disk model, circular orbits only exist at the equatorial plane. Any inclined bound photon orbits would acquire $\theta$-dependent deformations such that they do not have a constant radius. In Ref.~\cite{Chen:2022ynz}, it has been demonstrated that the eikonal correspondence of deformed Schwarzschild black hole spacetimes can be identified by defining the averaged radius of the bound photon orbits along one complete period. More explicitly, the averaged radius would correspond to the peak of the effective potentials of QNMs with $l\gg1$ and arbitrary $m_z$.

In this section, we will investigate the eikonal correspondence for the SBH-disk model. Specifically, we will consider the equatorial eikonal modes ($l=|m_z|\gg1$) and the polar eikonal modes ($m_z=0$ and $l\gg1$). The results obtained in this section can be treated as a consistency check with those exhibited in Ref.~\cite{Chen:2022ynz}.

\subsection{The equatorial modes $l=|m_z|$}
When $l=|m_z|$, the coefficients \eqref{coeffa}-\eqref{coeffd} can be expressed as
\begin{align}
a_{ll}^{2k}&=a_{l-l}^{2k}=\frac{l(2l+1)}{2}X_{\textrm{even}}(l,k)\,,\nonumber\\
b_{ll}^{2k}&=b_{l-l}^{2k}=\frac{2l+1}{2l+2k+1}X_{\textrm{even}}(l,k)\,,\nonumber\\
c_{ll}^{2k}&=c_{l-l}^{2k}=\frac{l(2l+1)(2k-1)}{2(2l+2k+1)}X_{\textrm{even}}(l,k)\,,\nonumber\\
d_{ll}^{2k}&=d_{l-l}^{2k}=-\frac{2kl(2l+1)}{2l+2k+1}X_{\textrm{even}}(l,k)\,,\nonumber\\
a_{ll}^{2k+1}&=a_{l-l}^{2k+1}=\frac{l(2l+1)}{2}X_{\textrm{odd}}(l,k)\,,\nonumber\\
b_{ll}^{2k+1}&=b_{l-l}^{2k+1}=\frac{2l+1}{2l+2k+1}X_{\textrm{odd}}(l,k)\,,\nonumber\\
c_{ll}^{2k+1}&=c_{l-l}^{2k+1}=\frac{l(2l+1)(2k-1)}{2(2l+2k+1)}X_{\textrm{odd}}(l,k)\,,\nonumber\\
d_{ll}^{2k+1}&=d_{l-l}^{2k+1}=-\frac{2kl(2l+1)}{2l+2k+1}X_{\textrm{odd}}(l,k)\,,
\end{align}  
where $k$ are non-negative integers, and
\begin{equation}
X_{\textrm{even}}(l,k)\equiv\frac{C^{l+k}_k}{C^{2l+2k}_{2k}}\,,\quad X_{\textrm{odd}}(l,k)\equiv\frac{C^{2l}_l}{4^lC^{l+k}_{k}}\,,
\end{equation}
where $C^i_j$ are the binomial coefficients. Therefore, in the eikonal limit $l\gg1$, only the coefficients $a_{ll}^0$ and $a_{l-l}^0$ dominate and read
\begin{equation}
a_{ll}^0=a_{l-l}^0\approx l^2\,.
\end{equation}
The effective potential \eqref{veff2} of the SBH-disk model can thus be approximated as
\begin{equation}
V_{\textrm{eff}}(r)\approx l^2\frac{f(r)}{r^2}\left[1+4\epsilon \mathcal{V}_0(r)\right]\,.\label{effeikonal1}
\end{equation}
In this case, the eikonal QNMs with $l=|m_z|\gg1$ correspond to the photons that undergo bound circular motion on the equatorial plane. According to Ref.~\cite{Chen:2022ynz}, the radius of these orbits is determined by the root of $\partial_r(g_{tt}/g_{\varphi\varphi})_{x=0}=0$. For the SBH-disk model with approximated metric \eqref{bhdiskmetric}, this equation can be written as
\begin{equation}
\partial_r\left[\frac{f(r)}{r^2}\left(1+4\nu_{\textrm{disk}}\right)\right]_{x=0}=0\,,\label{eqorbits}
\end{equation}
which is precisely the equation that determines the peak of the effective potential \eqref{effeikonal1} for $l=|m_z|\gg1$ because $\nu_\textrm{disk}|_{x=0}=\epsilon\mathcal{V}_0$.

\subsection{The polar modes $m_z=0$}

When $m_z=0$, the dominant coefficients in the eikonal limit $l\gg1$ are
\[
c_{l0}^{j}\approx
\begin{dcases}
    -\frac{1}{4^k}C^{2k}_kl^2\,,& \text{if } j=2k\\
    -\frac{4^{k+1}l^2}{\pi(k+1) C^{2k+2}_{k+1}}\,,              & \text{if } j=2k+1\,.
\end{dcases}
\]
The effective potential \eqref{veff2} is then approximated as
\begin{align}
V_{\textrm{eff}}(r)\approx l^2\frac{f(r)}{r^2}\Bigg\{1+\epsilon\sum_{k=0}^\infty\Bigg[\frac{C_{k}^{2k}}{4^k}\left(4\mathcal{V}_{2k}-2\mathcal{L}_{2k}\right)\nonumber\\+\frac{4^{k+1}}{\pi(k+1)C_{k+1}^{2k+2}}\left(4\mathcal{V}_{2k+1}-2\mathcal{L}_{2k+1}\right)\Bigg]\Bigg\}\,.\label{potentialpolarone}
\end{align}
Note that the last term on the right-hand side comes from the terms with odd $j$. 

The peak of the effective potential \eqref{potentialpolarone} can also be obtained through the calculations of photon geodesic equations. Consider the polar photon orbits on the photon sphere around the Schwarzschild black hole. Those orbits have zero azimuthal angular momentum $L_z$ and repeatedly reach the poles $x=\pm1$. When the spacetime is deformed, i.e., $\epsilon\ne0$, these polar orbits would also be deformed such that $\dot{r}=O(\epsilon)$ and $L_z=O(\epsilon)$, where the dot denotes the derivative along the geodesic with respect to the affine parameter $\lambda$. Up to the first-order of $\epsilon$, the radial component of the geodesic equations 
\begin{equation}
\frac{d}{d\lambda}\left(g_{\mu\nu}\dot{x}^\nu\right)=\frac{1}{2}\left(\partial_\mu g_{\alpha\beta}\right)\dot{x}^\alpha \dot{x}^{\beta}
\end{equation}
can be written as
\begin{equation}
\frac{d}{d\lambda}\left(g_{rr}\dot{r}\right)=\frac{1}{2}\frac{E^2}{g_{tt}}\partial_r\ln{\left|\frac{g_{tt}}{g_{\theta\theta}}\right|}+O\left(\epsilon^2\right)\,,
\label{poloargeodesicequation2}
\end{equation}
where $E$ is the energy of photons. Following Ref.~\cite{Chen:2022ynz}, we assume that the deformed orbits remain periodic and form a class of limit cycles in the phase space. We can then integrate Eq.~\eqref{poloargeodesicequation2} along a closed loop along $\lambda$. We then obtain
\begin{align}
o(\epsilon)&\propto\int_0^{2\pi}d\theta\partial_r\left(\frac{g_{tt}}{g_{\theta\theta}}\right)\nonumber\\
&\propto\int_0^{2\pi}d\theta\partial_r\left\{\frac{f(r)}{r^2}\left[1+\left(4\mathcal{V}_j(r)-2\mathcal{L}_j(r)\right)\left|\cos^j{\theta}\right|\right]\right\}\,,
\end{align}
with $j$ a dummy index standing for summations over all non-negative integers. Because of the absolute value of $\cos^j\theta$, both even and odd powers of $j$ contribute to the integration{\footnote{In Ref.~\cite{Chen:2022ynz}, the metric functions are expressed in series of $\cos\theta$ without absolute values. Therefore, in that case, only even powers of $j$ would contribute.}}. One can eventually get
\begin{align}
\partial_r&\Bigg\{\frac{f(r)}{r^2}\Bigg[1+\epsilon\sum_{k=0}^\infty\Bigg(\frac{C_{k}^{2k}}{4^k}\left(4\mathcal{V}_{2k}-2\mathcal{L}_{2k}\right)\nonumber\\&+\frac{4^{k+1}}{\pi\left(k+1\right)C_{k+1}^{2k+2}}\left(4\mathcal{V}_{2k+1}-2\mathcal{L}_{2k+1}\right)\Bigg)\Bigg]\Bigg\}=o(\epsilon)\,,\label{poloargeodesicequation3}
\end{align}
and then see that the root of Eq.~\eqref{poloargeodesicequation3} coincides with the peak of the effective potential \eqref{potentialpolarone}. In Ref.~\cite{Chen:2022ynz}, it has been proved that the root of Eq.~\eqref{poloargeodesicequation3} is precisely the averaged radius of the polar photon orbits along full periods. The averaged radius of bound photon orbits, both in the cases of equatorial orbits \eqref{eqorbits} and polar orbits \eqref{poloargeodesicequation3}, can be captured by their corresponding effective potentials of QNMs in the eikonal limit, i.e., Eqs.~\eqref{effeikonal1} and \eqref{potentialpolarone}, respectively. This is the manifestation of eikonal correspondence between bound photon orbits and high-frequency QNMs. Here, we show explicitly that even in the presence of spacetime deformations induced by a gravitating thin disk, as long as the disk mass is much smaller than the black hole mass, i.e., $\epsilon\ll1$, the eikonal correspondence can be identified through the definition of the averaged radius of bound photon orbits. This is consistent with the results of Ref.~\cite{Chen:2022ynz}.
 
\section{Conclusions}\label{sec.conclusion}

In this paper, we consider a recently obtained solution of deformed Schwarzschild black holes (SBH-disk model) \cite{Kotlarik:2022spo} and investigate the QNMs of a massless scalar field of this spacetime. The SBH-disk model describes the spacetime geometry of a Schwarzschild black hole encircled by a gravitating thin accretion disk. The superposed spacetime is an exact solution to GR and the gravitational field is regular everywhere outside the event horizon. In particular, the presence of the gravitating thin disk breaks the spherical symmetry, which is usually assumed in the literature when considering the gravitating fluid in the environment around astrophysical black holes.   

The lack of spherical symmetry of the SBH-disk model inevitably leads to the computational complexity of QNM frequencies because the angular and radial sectors of the QNM master equation are highly coupled. We overcome this difficulty by assuming that the disk mass $\mathcal{M}$ is much smaller than the black hole mass $M$. Up to the first order of $\mathcal{M}/M$, one can obtain the master equation that allows us to investigate the frequency shifts of QNMs in the presence of the disk. In particular, the radial sector of the master equation can be recast in a Schr\"odinger-like form in which the effective potential $V_\textrm{eff}(r)$ can be defined unambiguously and it reduces to the Schwarzschild one $V_\textrm{eff}^\textrm{Sch}(r)$ in proper limits.   

Besides the black hole mass $M$, the SBH-disk model contains four additional parameters, which essentially control the shape of the surface density profile for the disk. Taking a physically reasonable density profile, we find that the disk gravity would flatten the effective potential $V_\textrm{eff}(r)$ as compared with the Schwarzschild one. This behavior is robust among different choices of disk parameters. Furthermore, the presence of the gravitating disk would lower the real part of the QNM frequencies, while increase the damping time. In particular, the shifts of the real and imaginary parts with respect to their Schwarzschild counterparts, seem to follow a universal relation in the sense that they are shifted toward the same direction on the complex plane by the same amount in the presence of the disk (Fig.~\ref{fig.qnmuni}). Similar results also appear when the matter around the black hole is modeled based on the assumption of spherical symmetry \cite{Cardoso:2021wlq,Konoplya:2021ube}. Although still far away from a rigorous proof, if such a universal relation is indeed robust against the changes of matter configuration around the black hole, it would aid the discrimination between the disk effects on the QNM spectrum and those contributed by other putative physics beyond GR. This line of research deserves further investigation. 

In addition to QNM frequencies, we investigate two special kinds of bound photon orbits around the SBH-disk model. Since the equatorial symmetry is still preserved, the circular photon orbits on the equatorial plane exist, and the radius of the orbits is precisely at the peak of the effective potential of the eikonal equatorial QNMs. On the other hand, each polar orbit has a $\theta$-dependent radius because of the spacetime deformations. Assuming periodicity of the orbits, we find that the averaged radius of the orbits along a full period would correspond to the peak of the effective potential of the eikonal polar modes. This result is consistent with that found in the literature. 

In order to directly connect to the ringdown phase of gravitational waves, extending the present work to gravitational perturbations is necessary{\footnote{Similar analysis has been carried out in Ref.~\cite{Nagar:2006eu} in which the matter field around the black hole is assumed not to deform the black hole geometry at the background level, while interact gravitationally only at the perturbation level.}}. In addition, the physical properties of the SBH-disk model are not much explored so far. These include a detailed investigation of the geodesic dynamics of photons and massive particles. An extension towards a more realistic situation would be to include rotation of the black hole, the disk, or both in the black-hole--disk model and in the analysis of QNMs. We leave these interesting issues for future work. 

\acknowledgments
CYC is supported by the Institute of Physics of
Academia Sinica and the Special Postdoctoral Researcher (SPDR) Program at RIKEN. PK acknowledges support from GACR 21-11268S of the Czech Science Foundation.

\appendix

\section{Physical properties of the disks}\label{app.diskProperties}

To describe the matter content of infinitesimally thin disks, we have to introduce the stress-energy tensor on a singular hypersurface. Using the formalism developed by Israel \cite{Israel:1966}, the surface stress-energy tensor of a singular layer of matter located at $z=\text{const}$ reads \cite{Ledvinka:2019}
\begin{equation}
    S_{\alpha\beta} = -\frac{\sqrt{g_{\rho\rho}}}{8\pi} \left( \frac{g_{\alpha\beta}}{g_{\rho\rho}} \right)_{,z} \,.
    \label{wpSurfaceStressEnergy}
\end{equation}
This expression holds for any axially symmetric and stationary spacetime in Weyl coordinates. If the spacetime is static described by a metric \eqref{weylMetric}, the stress-energy tensor \eqref{wpSurfaceStressEnergy} has only two non-trivial components which read 
\begin{align}
    S_{tt} &= \frac{1}{2\pi} e^{3\nu - \lambda} \nu_{,z} (1 - \rho \nu_{,\rho}) \,, \\
    S_{\varphi\varphi} &= \frac{1}{2\pi} e^{-\lambda - \nu} \rho^3 \nu_{,z} \nu_{,\rho} \,,
\end{align}
where the right-hand sides are evaluated in the singular hypersurface, i.e., in the equatorial plane $z = 0$ where our disk lies. Consider a static observer equipped with a tetrad 
\begin{align}
    e_{(t)}^\alpha &= \frac{1}{\sqrt{-g_{tt}}} \delta^{\alpha}_t \,,  &e_{(\varphi)}^\alpha &= \frac{1}{\sqrt{g_{\varphi\varphi}}} \delta^{\alpha}_\varphi \,, \\
    e_{(\rho)}^\alpha &= \frac{1}{\sqrt{g_{\rho\rho}}} \delta^{\alpha}_\rho \,, &e_{(z)}^\alpha &= \frac{1}{\sqrt{g_{zz}}} \delta^{\alpha}_z \,.
\end{align}
In this tetrad, we easily observe that the disk can be interpreted as ideal fluid with density and azimuthal pressure (measured by the static observer hovering above the disk)
\begin{align}
    \sigma &\equiv S_{\alpha\beta}e^\alpha_{(t)} e^\beta_{(t)} = \frac{1}{2\pi} e^{\nu - \lambda} \nu_{,z} (1 - \rho \nu_{,\rho}) \,, \\
    P &\equiv S_{\alpha\beta} e^\alpha_{(\varphi)} e^\beta_{(\varphi)} = \frac{1}{2\pi} e^{\nu - \lambda} \rho \nu_{,z} \nu_{,\rho} \,.
    \label{physProp}
\end{align}

The relation between the $z$ derivative of the potential and the Newtonian surface density $w(\rho)$ can be obtained by integrating the Poisson equation $\Delta \nu = 4 \pi w(\rho) \delta(z)$ over the $z$ coordinate. Assuming that the spacetime is reflection symmetric with respect to the equatorial plane, only the term $\nu_{,zz}$ gives some non-zero contributions, thus
\begin{equation}
    w(\rho) = \frac{1}{2\pi} \lim_{z \rightarrow 0^+} \nu_{,z} \,.
\end{equation}
Substituting this relation into \eqref{physProp} we get precisely \eqref{physCharacteristics}.

\end{document}